\documentclass[11pt]{article}

\usepackage[utf8]{inputenc}
\usepackage[iso-8859-1]{inputenx}
\usepackage{graphicx}
\usepackage[english]{babel}
\usepackage{amsmath,mathtools}
\usepackage{amssymb}
\usepackage[margin=1in]{geometry}
\usepackage{hyperref}
\usepackage{float}
\usepackage{multirow}
\usepackage{latexsym}
\usepackage{caption}
\usepackage{babelbib}
\usepackage{latexsym}
\usepackage[T1]{fontenc}
\usepackage{tabularx}
\usepackage{dcolumn}
\usepackage{eurosym}
\usepackage{setspace}
\usepackage{color}
\usepackage[usenames,dvipsnames]{xcolor}
\usepackage{geometry}
\usepackage{multicol}
\usepackage{caption}
\usepackage{subfigure}
\usepackage{amsthm}
\usepackage{stmaryrd}
\usepackage{enumitem}
\usepackage{physics}
\usepackage{bbold}
\usepackage[bb=ams]{mathalfa}
\usepackage{bbm}
\usepackage{bbding}
\usepackage{wasysym}
\usepackage{amsmath,accents}

\usepackage[T1]{fontenc}
\usepackage{tgbonum}
\usepackage{lmodern}

\usepackage{amssymb}
\usepackage{amssymb}
\usepackage{pifont}
\usepackage{float}

\restylefloat{table}

\numberwithin{equation}{section}

\newtheorem*{theorem*}{Theorem}

\let\OLDthebibliography\thebibliography
\renewcommand\thebibliography[1]{
  \OLDthebibliography{#1}
  \setlength{\parskip}{1.5pt}
  \setlength{\itemsep}{2pt plus 2ex}
}

\title{\textsf{\textbf{Boundary conditions for Ashtekar-Barbero variables in the context of asymptotically flat spacetimes which lead to supertranslations at spatial infinity}}}
\author{
{\textrm{Sepideh Bakhoda}\thanks{s-bakhoda@bnu.edu.cn}}
\\
\\
{\rm \textit{Department of Physics, Beijing Normal University, Beijing 100875, China}}\\
}
\date{{\small\rm \today}}

\begin{document}

\maketitle
\sf{
\begin{abstract}
\rm{This paper delves into the exploration of suitable boundary conditions for the asymptotically flat scenario of general relativity presented in terms of Ashtekar-Barbero variables.  While the standard parity conditions have been extensively studied in \cite{Thiemann, Campiglia}, it turns out that  they fail to produce non-trivial supertranslations at spatial infinity. 
We propose new parity conditions for the Ashtekar-Barbero variables that do yield non-trivial supertranslation charges at spatial infinity. We compare our findings with those presented in \cite{Henneaux} and demonstrate that the new boundary conditions ensure the finiteness of the symplectic structure. Moreover, when embarking on the quest for appropriate parity conditions, it is essential to ensure that the selected parities remain invariant under hypersurface deformations. Given that working with Ashtekar-Barbero variables provides more asymptotic structure as compared to the ADM variables, it is shown that by fixing the Lagrange multiplier corresponding to the Gauss constraint, the invariance of certain parity conditions can be guaranteed.
}
\end{abstract}
}


\section{\textsf{Introduction}}
\rm{
In recent years, there has been a growing interest in the study of boundaries \cite{Han:2023nzl} in general relativity (GR) \cite{Wald:1984rg}, particularly with respect to the asymptotic symmetries of spacetimes. Bondi, van der Burg, and Metzner \cite{Bondi:1962px} discovered, and Sachs \cite{Sachs:1962wk} later confirmed, that the asymptotic region of an asymptotically Lorentzian flat spacetime has a richer structure than the traditional Poincar\'e group, which consists of translations, rotations, and boosts. The BMS group, a newly discovered symmetry group, is an extension of the finite-dimensional Poincar\'e group augmented by infinite-dimensional supertranslations \cite{Sachs:1962wk}.
Strictly speaking, the asymptotic region of an asymptotically flat spacetime can be viewed as an infinite set of classical vacua in general relativity. In this context, supertranslations describe the transformations that connect these vacua.

The BMS symmetry was initially observed at null infinity \cite{Bondi:1962px, Sachs:1962wk}. Nevertheless, studying the BMS symmetry at spatial infinity holds significant importance as well. There are four distinct motivations for analyzing the asymptotic structure of gravity at spatial infinity on spacelike hypersurfaces. 

First, supertranslations that are inextricably linked to the asymptotic structure of spacetime are accountable for the emission of gravitational radiation towards null infinity. Furthermore, it is well-known that gravitational radiations have the potential to devastate the typical smoothness requirements imposed at null infinity \cite{Christodoulou}. Hence, a fruitful approach to investigating the conditions on Cauchy data that yield a sufficiently smooth null infinity involves exploring the symmetries of the theory in a context that separates the presence of BMS symmetry from gravitational radiation.

Second, soft graviton theorems can be interpreted as Ward identities for the BMS asymptotic symmetries. Ward identities are mathematical equations that relate the symmetries of a system to its conserved charges \cite{Strominger:2017zoo}. However, constructing the BMS charges that generate the BMS symmetries in a canonical way is challenging.
One difficulty is that, at null infinity, it is more appropriate to consider fluxes rather than charges. Fluxes are measures of the flow of energy and momentum into or out of a system. Unfortunately, fluxes are not conserved when they are non-zero. Additionally, the BMS symmetries are not generated in a canonical way, and the association of functions with these symmetries is complex.
Hypersurfaces that reach null infinity are non-Cauchy, meaning that they do not capture the entire dynamical evolution of the system. Only when the fluxes at null infinity vanish can a standard Hamiltonian picture be recovered.
In conclusion, constructing the BMS charges that canonically generate the BMS symmetries is a challenging task due to the non-conservation of fluxes at null infinity and the non-canonical generation of the symmetries themselves.

Three,  pioneer studies \cite{RT, Thiemann, Campiglia} of the Hamiltonian structure at spatial infinity did not identify the BMS group as a group of physical symmetries, leading to a contradiction with results from null infinity. Therefore, it is important to resolve this tension in order to gain a better understanding of the symmetries of gravity.

Finally, our main motivation for studying BMS symmetries at spatial infinity lies in the quantum formulation of the theory, which is inherently explored on Cauchy hypersurfaces. 
The BMS algebra is manifested in the quantum theory through charges that act in the Hilbert space of states. These charges should have an expression at spatial infinity in the ADM formulation of evolution, which is based on foliations that approach asymptotic parallel hyperplanes, representing inertial observers at infinity.
Our ultimate objective is to examine the quantum properties of the symmetry and its associated charges within the framework of Loop Quantum Gravity (LQG) \cite{LQG}.

Since LQG is expressed in terms of the Ashtekar-Barbero variables  \cite{Ashtekar, BarberoG:1994eia}, the initial step towards achieving our goal is to determine appropriate boundary conditions for the canonical variables that allow for the emergence of BMS charges at spatial infinity. 
This has already been done in terms of the ADM variables in \cite{Henneaux}. In their approach, they have proposed a new boundary conditions compared to the boundary conditions in the earlier work \cite{RT}.
The boundary conditions employed in \cite{RT} at spatial infinity, with the intention of ensuring finite angular momentum, also result in all BMS charges becoming identically zero. This outcome, as demonstrated in \cite{RT}, arises due to the parity conditions imposed on the leading order of the metric and its conjugate momentum as one moves towards spatial infinity. Therefore, to reconcile the tension between the asymptotic structure at spatial infinity and the emergence of the BMS algebra at null infinity, it seems necessary to adopt boundary conditions at spatial infinity that differ from those of \cite{RT}.
Simply discarding the standard parity conditions is not feasible, as it leads to logarithmic divergence of the symplectic structure, angular momentum, and boost charges \cite{Beig:1987zz}. 
In the study by Henneaux and Troessaert \cite{Henneaux}, they discovered  alternative parity conditions for the leading terms of the metric and its conjugate momentum. These parity conditions preserve finiteness while allowing for a well-defined and nontrivial action of the BMS algebra. The key to their approach is to set the leading order terms of the constraints to zero. This is an additional restriction, but it is very mild since the leading terms of the constraints vanish on-shell anyway, and therefore do not remove any solutions.

In this paper, we investigate whether employing the same approach is possible to identify suitable boundary conditions for Ashtekar-Barbero variables, utilized in a LQG. The challenge lies in the increased asymptotic structure that must be determined due to the freedom in selecting the internal SU(2) frame, describing the internal orientation. However, to reproduce the ADM results from, it is necessary to fix the internal frame at the asymptotic boundary \cite{Thiemann, Campiglia}. This constraint is not inappropriate, as an SU(2) charge should not hold any physical significance in general relativity.
The optimal parity conditions simultaneously render the asymptotic symmetry generators and the symplectic 2-form finite while producing an integrable and finite charge. The fall-off and parity conditions of the additional degrees of freedom must be chosen in a manner that satisfies all these requirements concurrently. Furthermore, a challenging aspect of this work is ensuring that all imposed parities and fall-off conditions are preserved by the hypersurface deformation, which is a based task to tackle.
\\
\\
Our paper is structured as follows:
\\
\\
In Section 2, we begin by reviewing some classic background information on the ADM Hamiltonian treatment of asymptotically flat spacetimes, with particular emphasis on the parity conditions proposed by \cite{RT} and \cite{Henneaux}. We then proceed to discuss the Ashtekar-Barbero variables, deriving their fall-off conditions in asymptotically spherical coordinates. This is necessary because the new boundary conditions are most conveniently expressed in this coordinate system. Next, we recall the standard boundary conditions for the Ashtekar-Barbero variables obtained in \cite{Thiemann} and \cite{Campiglia}, as well as the strategy employed by these papers to fix the internal frame at the asymptotic boundary.
\\
In Section 3, we begin by stating a theorem that introduces fall-off and parity conditions for the canonical variables and expresses and fixes some of the leading terms of the Lagrange multipliers. We then spend the remainder of the section proving the theorem. Specifically, we show that the symplectic form is finite and well-defined, the boundary conditions are preserved by hypersurface deformations, and the constraints are well-defined and functionally differentiable for asymptotic translations and do yield to non-trivial supertranslations at spatial infinity. Thus, we derive a finite and integrable charge for infinite-dimensional supertranslations.
\\
In Section 4, an analysis is carried out to assess the correlation between our findings and the outcomes presented in \cite{Henneaux}, specifically regarding the retrieval of the supertranslation charge within the ADM formulation. Furthermore, a  discussion is provided on the distinction observed in the boundary terms when employing ADM variables in comparison to Ashtekar-Barbero variables.
\\
In the final section, we summarize our findings and provide an outlook. The conclusive outcomes of some lengthy calculations have been included in an appendix.

\section{\textsf{Background}}
This section is dedicated to establishing notation and outlining the necessary steps for verifying the suitability of boundary conditions. Additionally, a brief review of previous work will be provided.

In this paper, we employ the Hamiltonian formalism of GR. Consequently, it is important to emphasize that this formulation assumes a foliation of spacetime into spacelike hypersurfaces. Consistent with the canonical approach, we select a Cauchy hypersurface $\Sigma$ and define cartesian coordinates $x^a = (x, y, z)$ on it. The region at spatial infinity is identified as $r \to \infty$, where $r = x^a x_a$.

Appropriate boundary conditions are those that satisfy the following requirements collectively. These requirements will be revisited throughout the paper. Consistent boundary conditions are those that:

\begin{enumerate}
\item Ensure the symplectic structure is well-defined.\\
A well-defined symplectic structure is crucial for working in the phase space and calculating Poisson brackets. Therefore, it is necessary to ensure that the proposed boundary conditions do not cause the symplectic structure to become divergent as one approaches infinity.
\item Remain invariant under hypersurface deformations.\\
To establish a consistent theory, it is essential for the boundary conditions to remain the same across all slices. In other words, they must be invariant as one moves from one hypersurface to the next.
\item Enable the Hamiltonian generators of the asymptotic symmetries to be well-defined and integrable.\\
The surface integrals that yield the charges associated with the asymptotic symmetries should be finite and integrable. By integrable, we mean that the variation can be extracted from the integral in the surface integral. Mathematically speaking, the variation of the surface charge is a one-form in field space obtained by performing integration by parts on the bulk generator. This one-form must be exact.\\
\\
This requirement can be explained in another way: \\
In gauge theories, including GR, one has to work with some constraints whose vanishing represents equations relating the canonical variables.
The constraint surface of the phase space is defined by the vanishing of these constraint functionals. 
Should these constraints meet the first-class criterion in Dirac's terminology \cite{Henneaux:1992ig}, they are responsible for generating gauge transformations. 
Consequently, it becomes necessary to calculate Poisson brackets with these constraint functionals.
In order to compute Poisson-brackets between the constraint functionals and different functions on the phase space, it is essential for them to be both finite and functionally differentiable.

If, in the presence of a boundary, it occurs that the generators of gauge transformations are not functionally differentiable, one can address this issue by following the subsequent procedure:
Compute the variation of the constraint, which involves obtaining a surface integral. If the variation's volume term is well-defined, it yields the desired functional derivative of the functional that we aim to establish as well-defined. Subtract the surface term from the variation of the original constraint.
If, after this process, the resulting surface term is found to be exact, meaning it can be expressed as the variation of a surface integral, then one has acquired an expression that is functionally differentiable and, if fortunate, is already finite and so it is well-defined.
\end{enumerate}

In a totally constrained system, the calculation of the charge requires the consideration of the variation of the sum of the smeared constraints and the identification of appropriate surface terms that can be added to the sum to ensure differentiability. In mathematical terms, let us consider a fully constrained system characterized by canonical variables $(q^a, p_a)$ and first-class constraints $C_I$ with corresponding smearing functions $\lambda^I$. The variation can then be expressed as
\begin{equation}\label{def. surface term}
\delta C_I [\lambda^I] = \int d^3x \; (\delta_{\lambda} p_a \delta q^a - \delta_{\lambda} q^a \delta p_a) +\mathcal{B}_{\lambda} (\delta q^a, \delta p_a)
\end{equation} 
Here, $\mathcal{B}_{\lambda} [\delta q^a, \delta p_a]$ represents the surface term. Initially, it is crucial to ensure that $\mathcal{B}_{\lambda} [\delta q^a, \delta p_a]$ converges as the boundary is approached. Subsequently, it must be verified if this term can be expressed as the variation of a surface integral. If both of these criteria are met, the charge is determined by $\mathcal{Q}_\lambda (q^a, p_a)$, defined as 
\begin{equation}\label{definition of charge}
\mathcal{B}_{\lambda} (\delta q^a, \delta p_a) = - \delta \mathcal{Q}_\lambda (q^a, p_a).
\end{equation}



\subsection{\textsf{Boundary conditions for ADM variables}}

Within the framework of the ADM formalism, the spatial metric tensor $q_{ab}$ of $\Sigma$, as well as the lapse function $N$ and the shift vector $N^a$, along with their conjugate momenta $\pi^{ab}$, $\Pi$, and $\Pi_a$, respectively, serve as the canonical variables. To simplify notation, we define $\textbf{N}:=(N, N^a) = (N, \vec{N})$. Upon deriving the ADM action, it becomes apparent that the action is independent of the time derivatives of $N$ and $N^a$. Consequently, this leads to the establishment of primary constraints $\Pi =0$ and $\Pi_a =0$. Regarding the conjugate momentum of the metric, it can be determined that $\pi^{ab} = \sqrt{q} (K^{ab}-K q^{ab})$, where $q$ denotes the determinant of the spatial metric $q_{ab}$ and $K_{ab} = \frac{1}{2N}(\dot{q}_{ab}-\mathcal{L}_{\vec{N}}q_{ab} )$ represents the extrinsic curvature of $\Sigma$ with $K$ representing its trace.

Stability of the primary constraints shows that the secondary constraints are (for a complete review of the geometrodynamics of GR the reader refer to \cite{ThiemannBook})
\begin{align}
H_a [N^a]&:= -2\int_{\Sigma} d^3x \; N^a q_{ac} D_b P^{bc} \label{diff. Const. ADM}\\
H [N] &:= -  \int_{\Sigma} d^3x \; N \left(\frac{s}{\sqrt{q}} \left[q_{ac} q_{bd} - \frac{1}{2}q_{ab} q_{cd} \right] \pi^{ab} \pi^{cd} + \sqrt{q} R \right) \label{Ham. Const. ADM}
\end{align}
where $R$ is the Ricci scalar of the spatial hypersurface $\Sigma$, and $D_a$ is the Levi-civita connection associated to $q_{ab}$. By imposing stability of these constraints under evolution, no tertiary constraints arise. 
The complete introduction of the phase space of the theory is achieved through the definition of the symplectic structure. Denoted as $\Omega(\delta_1, \delta_2)$, this symplectic structure is given by 
\begin{equation}\label{Symplectic Struc. of ADM}
\Omega(\delta_1, \delta_2) = \int_\Sigma d^3x\; (\delta_1 q_{ab} \delta_2 \pi^{ab}-\delta_2 q_{ab} \delta_1 \pi^{ab})
\end{equation}
The variation of a phase space function $\mathcal{F}[q, \pi]$ is defined as $\delta \mathcal{F} = \Omega(\delta_{\mathcal{F}}, \delta)$ where $\delta =(\delta q_{ab}, \delta \pi^{ab})$ and $\delta_{\mathcal{F}} =(\delta_{\mathcal{F}} q_{ab}, \delta_{\mathcal{F}} \pi^{ab})$ and $\delta_{\mathcal{F}}$ is simply the Hamiltonian vector field of $\mathcal{F}$. Then the Poisson bracket of two given space time functions $\mathcal{F}_1$ and $\mathcal{F}_2$ is defined by $\{\mathcal{F}_1, \mathcal{F}_2\} := \Omega(\delta_{\mathcal{F}_1}, \delta_{\mathcal{F}_2})$ \cite{Campiglia}.

\subsubsection{\textsf{R-T boundary conditions}}\label{Section R-T boundary conditions}
After providing this brief overview, we are now prepared to explore the examination of the asymptotic region using the canonical formulation of GR. 
A spacetime is considered asymptotically flat\footnote{Mathematically rigorous definitions for an asymptotically flat spacetime exist, which are beyond the scope of this paper \cite{DeWitt:1984ojp}.} if, outside of a compact region, the metric follows the behavior $g_{\mu \nu} = \delta_{\mu \nu} + \frac{1}{r} h_{\mu \nu}$, where $h_{\mu \nu}$ is a tensor on the asymptotic 2-sphere ($\partial \Sigma = S^2$). In order to make use of the Hamiltonian formalism, it is essential to comprehend the decay behaviors of the variables $q_{ab}$ and $\pi^{ab}$. Although there is no indication of the decay behavior of the latter in the fall-off behavior of $g_{\mu \nu}$, the former can be directly derived from it. 
Demanding the symplectic structure (\ref{Symplectic Struc. of ADM}) to be finite, one arrives at this conclusion that $\pi^{ab} = O(r^{2+\epsilon})$ (for more detail look at \cite{RT} or the second reference in \cite{LQG}). However, it is precisely the finiteness of the ADM momentum that necessitates the decay of $r^{-2}$. Therefore, one must adhere to the fall-off behaviors as
\begin{equation} \label{BC Cartesian}
    \begin{split}
q_{ab} &= \delta_{ab}+\frac{1}{r} \bar{h}_{ab} (\vec{\mathbf{n}}) + O(r^{-2}) \\
\pi^{ab} &= \frac{1}{r^2} \bar{\pi}^{ab}(\vec{\mathbf{n}}) + O(r^{-3})
    \end{split}
\end{equation}
Here, $\bar{h}_{ab}$ and $\bar{\pi}^{ab}$ are tensor fields on the 2-sphere at spatial infinity ($\vec{\mathbf{n}}=\frac{\vec{x}}{r}$), and now the objective is to eliminate the divergences arising in (\ref{Symplectic Struc. of ADM}) through an alternate technique.  This is exactly where the parity conditions come into play. A possible way to eliminate the divergence in (\ref{Symplectic Struc. of ADM}) is to impose the condition that the functions $\bar{h}_{ab}$ and $\bar{\pi}^{ab}$ possess opposite parity. In \cite{RT}, the following parity conditions were proposed, which are commonly known as the \textit{standard parity conditions}. Under the antipodal map on the asymptotic 2-sphere, $\bar{h}_{ab}$ and $\bar{\pi}^{ab}$ show the following behavior
\begin{equation}\label{Standard parity conditions (ADM)}
\bar{h}_{ab} \left(- \vec{\mathbf{n}}\right)= \bar{h}_{ab} \left(\vec{\mathbf{n}}\right), \;\;\;\;\;\; \bar{\pi}^{ab} \left(- \vec{\mathbf{n}}\right) = - \bar{\pi}^{ab} \left(\vec{\mathbf{n}}\right)
\end{equation}
In other words, $\bar{h}_{ab}=\text{even}$ and $\bar{\pi}^{ab}=\text{odd}$. 
Looking at the power expansion of the symplectic 2-form (\ref{Symplectic Struc. of ADM}), i.e.
\begin{equation}
\Omega(\delta_1, \delta_2) = \int \frac{dr}{r} \int_{S^2} d\mathbf{\sigma} \; (\delta_1 \bar{h}_{ab} \delta_2 \bar{\pi}^{ab} - \delta_2 \bar{h}_{ab} \delta_1 \bar{\pi}^{ab}) + \text{finite}
\end{equation}
one finds that using the parity conditions (\ref{Standard parity conditions (ADM)}) the coefficient of the leading logarithmic singularity is zero because the term $\delta \bar{h}_{ab} \delta \bar{\pi}^{ab}$  is an odd function and its integral over the sphere vanishes. 
Here, $d\mathbf{\sigma}$ is the standard measure on the unit sphere.
It is worth mentioning that it is not possible to interchange the parity conditions (\ref{Standard parity conditions (ADM)}) due to the resulting in disappearance of the ADM energy momentum.

Now in accordance with (iii), one needs to verify if the constraints (\ref{diff. Const. ADM}) and (\ref{Ham. Const. ADM}) are well defined. To accomplish this, the asymptotic behaviors of the Lagrange multipliers, i.e., $N$ and $N^a$, must first be determined. In the simplest version, in [4] it is shown that the Hamiltonian and diffeomorphism constraints are finite and functionally differentiable when the lapse and shift have the following $r \to \infty$ asymptotic behavior
\begin{equation} \label{Simple version of Lapse and Shift}
N= S(\vec{\mathbf{n}}) + O(r^{-1}), \;\;\;\;\;\; N^a = S^a(\vec{\mathbf{n}})+O(r^{-1})
\end{equation}
where $S(\vec{\mathbf{n}})$ and $S^a(\vec{\mathbf{n}})$ are arbitrary odd functions on the unit sphere, i.e.
\begin{equation}\label{Parity of Simple version of Lapse and Shift}
S(- \vec{\mathbf{n}}) = - S(\vec{\mathbf{n}}), \;\;\;\;\;\; S^a(-\vec{\mathbf{n}}) = - S^a(\vec{\mathbf{n}})
\end{equation}
Thus, the constraints (\ref{diff. Const. ADM}) and (\ref{Ham. Const. ADM}) with lapse and shift obeying (\ref{Simple version of Lapse and Shift}), (\ref{Parity of Simple version of Lapse and Shift}) generate the gauge transformation of the theory. Note that the charge corresponding to the so-called supertranslation $S(\vec{\mathbf{n}})$ and $S^a(\vec{\mathbf{n}})$ is identically zero. We will return to this point in section \ref{Section Asymptotic charges}.

In the context of asymptotically flat spacetime, it is reasonable to allow for the decay behaviors of smearing functions $N, N^a$ which correspond to infinitesimal Poincar\'e transformations. These behaviors can be described as follows
\begin{align}\label{lapse-shift Cartesian}
N= b_a x^a + a(\vec{\mathbf{n}}) + O(r^{-1}), \;\;\;\;\;\; N^a = b^a_b x^b + a^a(\vec{\mathbf{n}})+O(r^{-1})
\end{align}
where $b_a$ and $b_{ab} = -b_{ba}$ represent arbitrary constants, while $a(\vec{\mathbf{n}})$ and $a^a(\vec{\mathbf{n}})$ are arbitrary functions on the unit sphere. The constants $b_a$ serve as parameters for Lorentz boosts, and the antisymmetric constants $b_{ab} = -b_{ba}$ act as parameters for spatial rotations. Arbitrary functions $a$ and $a^a$ describe angle-dependent translations whose zero modes are standard translations.

In their paper \cite{RT}, Regge and Teitelboim demonstrated that the boundary conditions (\ref{BC Cartesian}) and (\ref{Standard parity conditions (ADM)}) not only ensure the finiteness of the symplectic structure (\ref{Symplectic Struc. of ADM}), but also possess invariance under hypersurface deformations with lapse and shift (\ref{lapse-shift Cartesian}) provided $a, a^a$ (except their zero modes) are odd functions, thereby satisfying conditions (i) and (ii). However, when introducing the Lagrange multipliers as (\ref{lapse-shift Cartesian}), it turns that the constraints (\ref{diff. Const. ADM}) and (\ref{Ham. Const. ADM}) are not well-defined. Following the procedure outlined in (iii), the authors were able to identify well-defined generators. Hence, it can be concluded that the boundary conditions (\ref{BC Cartesian}) and (\ref{Standard parity conditions (ADM)}) are appropriate as they fulfill all the requirements (i)-(iii). The only limitation of their work is that the resulting charge only includes those associated with Poincar\'e symmetries, leaving no scope for charges associated with the non-constant angle-dependent translations $a, a^a$. Therefore, there is no room for the BMS charge.
To address this issue, an alternative set of appropriate boundary conditions must be sought. This is precisely what Henneaux and Troessaert accomplished in their study \cite{Henneaux}. Below, we briefly outline their approach and encourage interested readers to consult the original paper for further details.

\subsubsection{\textsf{H-T boundary conditions}}\label{Section H-T boundary conditions}
In order to investigate the asymptotic region, it is preferable to use spherical coordinates $(r, x^A)$, where $x^A$ denotes the coordinates on the 2-sphere. In what follows depending on the definition of the antipodal map on the 2-sphere, two types of coordinates are utilized
\begin{itemize}
\item[1.] The coordinates for which the antipodal map is $x^A \to - x^A$. \label{Coordinate 1}
\item[2.] The traditional coordinates $x^A =(\theta, \varphi)$ for which the antipodal map is $\theta \to \pi - \theta, \varphi \to \varphi + \pi$. \label{Coordinate 2}
\end{itemize}
When working with tensorial equations, the choice of coordinates does not affect the results. However, in terms of associating parities, it is important to specify the coordinate system being used. It is always possible to perform a coordinate transformation to switch between these two coordinate systems. 
In a slight abuse of notation, we utilize $x^A$ to represent both sets of coordinates (1) and (2). However, if any confusion arises, we will explicitly clarify which coordinate system is being used. We will refer to them as coordinate system (1) and coordinate system (2) respectively.

When expressed in the spherical coordinates, equation (\ref{BC Cartesian}) can be written as
\begin{equation}\label{BC-Spherical}
     \begin{split}
         & q_{rr} = 1 + \frac{1}{r}\bar{h}_{rr} + \frac{1}{r^2} h^{(2)}_{rr} + O(r^{-3})\\
         & q_{rA} = \bar{h}_{rA} + \frac{1}{r}h^{(2)}_{rA} + O(r^{-2})\\
         & q_{AB} = r^2 \bar{\gamma}_{AB} + r\bar{h}_{AB} + h^{(2)}_{AB} + O(r^{-1})\\
         & \pi^{rr} = \bar{\pi}^{rr} + \frac{1}{r}\pi^{(2)rr} + O(r^{-2})\\
         & \pi^{rA} = \frac{1}{r}\bar{\pi}^{rA} + \frac{1}{r^2} \pi^{(2)rA} + O(r^{-3})\\
         & \pi^{AB} = \frac{1}{r^2}\bar{\pi}^{AB} + \frac{1}{r^3}\pi^{(2)AB} + O(r^{-4}),
     \end{split}
 \end{equation}
 where $\bar{\gamma}_{AB}$ is the metric on the unit 2-sphere. It should be noted that when deriving the last three equations in (\ref{BC-Spherical}), one must consider that $\pi^{ab}$ is not a tensor field but rather a tensor density. 
 Furthermore, it is possible to assume, without loss of generality, that $\bar{h}_{rA}=0$, which greatly simplifies the calculations in subsequent sections. This assumption holds true since $\bar{h}_{rA}=0$ can always be achieved through a coordinate transformation \cite{Henneaux}.

Expressed in spherical coordinates, equation (\ref{lapse-shift Cartesian}) takes the form:
\begin{align}
        &N = r b+f + O(r^{-1}), \label{lapse Spherical}\\
         N^r = W  +& O(r^{-1}), \;\; N^A = Y^A + \frac{1}{r}I^A + O(r^{-2}) \label{shift Spherical}
\end{align}
Here, $b$, an arbitrary function satisfying the condition $\bar{D}_A \bar{D}_B b + b \bar{\gamma}_{AB} = 0$, serves as the boost parameter, and $Y^A$ is assumed to be the rotation generator, satisfying $\mathcal{L}_Y\bar{\gamma}_{AB} = 0$. It should be noted that $\bar{D}_A$ denotes the torsion-free connection that is compatible with the metric of the unit 2-sphere $\bar{\gamma}_{AB}$.

When attempting to compute the surface term $\mathcal{B}_\lambda [\delta q^a, \delta p_a]$ defined in (\ref{def. surface term}) for the case of GR with the constraints given by  (\ref{diff. Const. ADM}) and (\ref{Ham. Const. ADM}), and with smearing functions defined in (\ref{lapse Spherical}) and (\ref{shift Spherical}), it is found that  $\mathcal{B}_{\mathbf{N}} [\delta q_{ab}, \delta \pi^{ab}]$ exhibits linear divergence, specifically
\begin{equation}\label{div. term of B ADM}
 \left(\mathcal{B}_{\mathbf{N}} [\delta q_{ab}, \delta \pi^{ab}]\right)_{\text{divergent part}} = r \int_{S^2} d\sigma \; (-2Y^A\bar{\gamma}_{AB}\delta\bar{\pi}^{rB} - 2\sqrt{\bar{\gamma}}b\delta\bar{k})
 \end{equation}
where
\begin{equation}\label{def. K and lambda}
\bar{k}:= \bar{\gamma}^{AB} \bar{k}_{AB}, \;\;\;\; \bar{k}_{AB} := \frac{1}{2} \bar{h}_{AB} + \bar{\lambda} \bar{\gamma}_{AB},\;\;\;\; \bar{\lambda} := \frac{1}{2} \bar{h}_{rr}
\end{equation}
In the work by Henneaux and Troessaert \cite{Henneaux}, it was observed that the divergent term (\ref{div. term of B ADM}) can be eliminated with the help of the leading order terms of the constraints, i.e. $\bar{H}, \bar{H}_A$ where $H=: \frac{1}{r} \bar{H}+ O(r^{-2})$ and $H_A =: \bar{H}_A + O(r^{-1})$. Specifically, the conditions $\bar{H}=0=\bar{H}_A$ were imposed even off-shell, where
\begin{equation}\label{LOT of constraints ADM}
    \begin{split}
        & \bar{H}=\bar{D}_A\bar{D}_B \bar{k}^{AB} - \bar{D}_A\bar{D}^A \bar{k} = 0, \;\;\; \bar{H}_A= \bar{\gamma}_{AB}\bar{\pi}^{rB} + \bar{\gamma}_{AB}\bar{D}_C\bar{\pi}^{BC} = 0.\\
    \end{split}
\end{equation}
By utilizing equation (\ref{LOT of constraints ADM}), performing integration by parts, and employing the boost property $\bar{D}_A\bar{D}_B b + b\bar{\gamma}_{AB} = 0$, the divergent term (\ref{div. term of B ADM}) can be eliminated. In the previous work by Regge and Teitelboim \cite{RT}, the removal of (\ref{div. term of B ADM}) was achieved through parity conditions. However, in the study conducted by Henneaux and Troessaert \cite{Henneaux}, the role is fulfilled by (\ref{LOT of constraints ADM}), enabling the relaxation of parity conditions and the introduction of different conditions than those proposed by R-T, thus leading to the revival of the supertranslation charge.

To present the H-T parity conditions, besides the newly defined variables (\ref{def. K and lambda}) we need to define
\begin{equation}\label{def. bar p}
\bar{p} := 2(\bar{\pi}^{rr}-\bar{\pi}^A_A)
\end{equation}
Then in terms of the spherical coordinates (1), the set of parity conditions on the boundary values proposed in \cite{Henneaux} are
\begin{equation}\label{H-T-Parity 1}
    \begin{split}
        \bar{\lambda} \sim \bar{\pi}^{AB} = even, \;\;\;\;\; \bar{p} \sim \bar{k}_{AB} \sim \bar{\pi}^{rA} = odd, 
    \end{split}
\end{equation}
or in terms of the spherical coordinates (2),
\begin{equation}\label{H-T-Parity 2}
    \begin{split}
        &\bar{\lambda} \sim  \bar{\pi}^{r\varphi}  \sim \bar{\pi}^{\theta \theta} \sim \bar{\pi}^{\varphi \varphi} \sim \bar{k}_{\theta \varphi}  = \text{even}, \\
        &\bar{p} \sim  \bar{\pi}^{r\theta}  \sim \bar{\pi}^{\theta \varphi} \sim \bar{k}_{\theta \theta} \sim \bar{k}_{\varphi \varphi}  = \text{odd}.
    \end{split}
\end{equation}
Thus, (\ref{BC-Spherical}) together with (\ref{H-T-Parity 1}) are called the H-T boundary conditions \cite{Henneaux}.
It is straightforward to see that these boundary conditions remove the divergent term of the symplectic structure, because
\begin{equation}
\begin{split}
\Omega(\delta_1, \delta_2) &= \int \frac{dr}{r} \int_{S^2} d\mathbf{\sigma} \; \left(\delta_1 \bar{h}_{rr} \delta_2 \bar{\pi}^{rr} + \delta_1 \bar{h}_{AB} \delta_2 \bar{\pi}^{AB} - (\delta_1 \leftrightarrow \delta_2)\right) + \text{finite}\\
&=
\int \frac{dr}{r} \int_{S^2} d\mathbf{\sigma} \; \left(\delta_1 \bar{h}_{rr} \delta_2 \bar{\pi}^{rr} + \delta_1 \bar{h}_{AB} \delta_2 \bar{\pi}^{AB} +\delta_1 \bar{h}_{rr} \delta_2 \bar{\pi}^A_A - \delta_1 \bar{h}_{rr} \delta_2 \bar{\pi}^A_A - (\delta_1 \leftrightarrow \delta_2)\right) + \text{finite}\\
&=
\int \frac{dr}{r} \int_{S^2} d\mathbf{\sigma} \; \left(\delta_1 \bar{h}_{rr} \delta_2 [\bar{\pi}^{rr} -\bar{\pi}^A_A] + \delta_1 [\bar{h}_{AB} + \bar{h}_{rr} \bar{\gamma}_{AB}] \delta_2 \bar{\pi}^{AB}   - (\delta_1 \leftrightarrow \delta_2)\right) + \text{finite}\\
&=
\int \frac{dr}{r} \int_{S^2} d\mathbf{\sigma} \; \left(\delta_1 \bar{\lambda} \delta_2 \bar{p} + 2 \delta_1 \bar{k}_{AB}  \delta_2 \bar{\pi}^{AB}   - (\delta_1 \leftrightarrow \delta_2)\right) + \text{finite}
\end{split}
\end{equation}
and in accordance with (\ref{H-T-Parity 1}), the terms $\delta \bar{\lambda} \delta \bar{p}$ and $\delta \bar{k}_{AB}  \delta \bar{\pi}^{AB} $  are odd functions and their integral over the sphere are equal to zero. It should be noted that the term $\delta \bar{h}_{rA} \delta \bar{\pi}^{rA}$ does not contribute to the divergent term of the symplectic structure due to the assumption $\bar{h}_{rA}=0$. Through a lengthy calculation, it can be demonstrated that the H-T boundary conditions remain invariant under hypersurface deformations, thus meeting the requirements (i) and (ii).  As previously mentioned, the divergent part of the surface term (\ref{div. term of B ADM}) is eliminated by setting the leading terms of the constraints to zero, i.e. (\ref{LOT of constraints ADM}). Furthermore, it has been proven that the surface term $\mathcal{B}_{\mathbf{N}} [\delta q_{ab}, \delta \pi^{ab}]$ is exact, thereby establishing the integrability of the charge  \cite{Henneaux}. Therefore, all the requirements (i)-(iii) are satisfied by H-T boundary conditions (\ref{Standard parity conditions (ADM)}) and (\ref{H-T-Parity 1}).

The non-zero supertranslation charge is hence given by 
\begin{equation}\label{STcharge}
    \begin{split}
        \mathcal{Q}_{Supertranslation} = \int_{S^2} d\sigma \; ( 4T\sqrt{\bar{\gamma}}\bar{\lambda} + W \bar{p}),
    \end{split}
\end{equation}
where $T:=f+b\bar{\lambda}+b\bar{k}$ and $W$ are even and odd arbitrary functions on the unit 2-sphere, respectively. The parities of $T$ and $W$ ensure that the H-T parity conditions (\ref{H-T-Parity 1}) remain invariant under hypersurface deformations. Note that the terms given in (\ref{STcharge}) may not generally vanish as they are determined by integrating arbitrary even functions.

It is noteworthy to examine why the R-T parity conditions result in a vanishing supertranslation charge. 
As mentioned in the concluding paragraph of section \ref{Section R-T boundary conditions}, the non-zero modes of the arbitrary functions $a$ and $a^a$ are odd. This can be equivalently expressed in terms of spherical coordinates, indicating that the non-zero modes of $f$ and $W$ are odd and even, respectively. Furthermore, when the R-T parity conditions (\ref{Standard parity conditions (ADM)}) are translated into spherical coordinates, it becomes evident that $\bar{\lambda} =\text{even}$ and $\bar{p}\sim \bar{k}_{AB} =\text{odd}$, leading to even parity for the non-zero mode of $T$. Considering all of these factors, we can arrive at the conclusion that the arbitrary functions present in $T$ and $W$ have opposite parity to $\bar{\lambda}$ and $\bar{p}$, respectively. Equation (\ref{STcharge}) indicates that these functions possess identically vanishing surface charges, leaving no room for the BMS symmetry with the parity conditions (\ref{Standard parity conditions (ADM)}).

\subsection{\textsf{Standard boundary conditions for Ashtekar-Barbero variables}}\label{Section Standard BC Ash.}
The Ashtekar-Barbero formalism relies on the use of tetrad variables to represent the gravitational field. 
These variables consist of four covariant fields denoted as $e^I_\mu (x)$, where $I, J, \cdots = 0, 1, 2, 3$ are flat indices that are raised and lowered by the metric $\eta_{IJ} = \text{diag}[s, +1, +1, +1]$. Here, $s$ represents the signature of the metric ($s=+1$ for Euclidean and $s=-1$ for Lorentzian metric). The metric variables can be expressed in terms of the tetrad variables using the equation $g_{\mu \nu} = \eta_{IJ}e^I_\mu e^J_\nu$. This formulation introduces an additional gauge invariance of $SO(3,1)$ into GR. 
The corresponding canonical formalism is defined within the temporal gauge $e^i_t = 0$, where $i, j, \cdots = 1, 2, 3$ are flat three-dimensional indices raised and lowered by $\delta_{ij}$. In this gauge, the Lorentz group is reduced to $SU(2)$, and the ADM configuration variables are $q_{ab}=e^i_a e^i_b, N= e^0_t, e^i_t = e^i_a N^a$.

The Ashtekar-Barbero variables \cite{Ashtekar, BarberoG:1994eia} consist of the following connection 1-form and electric field
 \begin{align}
A^i_a &:= \Gamma^i_a + \beta K^i_a \label{Ashtekar connection}\\
E^a_i &:= \sqrt{q} \; e^a_i \label{Electric field Ashtekar}
\end{align}
respectively, where $q:= \det (q_{ab})$,  $ e^a_i$ is defined by the relations $e^a_i e^j_a= \delta^i_j, \;e^a_i e^i_b = \delta^a_b$, the one-form $K^i_a$ is defined through $K_{ab}=: K^i_{(a} e^i_{b)}$, the parameter $\beta$ is an arbitrary complex number known as Barbero-Immirzi parameter \cite{BarberoG:1994eia} and
\begin{equation}\label{Spin connection}
\begin{split}
\Gamma^i_a :=& -\frac{1}{2} \epsilon_{ijk} E^j_b D_a E^b_k \\
=& -\frac{1}{2} \epsilon^{ijk} E^b_k \left[E^j_{a,b}-E^j_{b,a}+ E^c_j E^l_a E^l_{c,b} + E^j_a \frac{(\det(E))_{,b}}{\det(E)} \right]
\end{split}
\end{equation}
is the spin connection associated with $e^i_a$ in which $E^i_a$ is the inverse of (\ref{Electric field Ashtekar}). The phase space coordinated by the pair $(A^i_a, E^a_i)$ is equipped with the symplectic structure given by
\begin{equation}\label{Symplectic Struc. of Ash.}
\Omega(\delta_1, \delta_2) = \frac{2}{\beta} \int_\Sigma d^3x\; (\delta_1 A^i_a \delta_2 E^a_i -\delta_2 A^i_a \delta_1 E^a_i)
\end{equation}
with respect to which the Ashtekar-Barbero variables (\ref{Ashtekar connection}) and (\ref{Electric field Ashtekar}) form a canonically conjugate pair. 
In terms of these variables, the constraints of the theory are expressed in the form \cite{ThiemannBook}
\begin{align}
&\mathcal{G}_i [\Lambda^i] = \frac{2}{\beta} \int_\Sigma d^3x\; \Lambda^i \left(\partial_a E^a_i + \epsilon_{ijk}A^j_a E^a_k\right) \label{Gauss Cons.}\\
&H_a [N^a] =\frac{-2 s}{\beta} \int_\Sigma d^3x\; N^a \left(F^i_{ab}E^b_i - A^i_a \mathcal{G}_i \right) \label{diff. Cons. Ash}\\
&H [N] = \int_\Sigma d^3x\; \tilde{N} \left[F_{ab}^i -(\beta^2-s)\epsilon_{imn} K^m_a K^n_b \right] \epsilon_{ijk}E^a_j E^b_k \label{Ham. Cons. Ash}
\end{align}
that are known as Gauss, diffeomorphism and Hamiltonian constraints, respectively. Here,
\begin{equation}\label{Fab}
F^i_{ab}=\partial_aA^i_b-\partial_bA^i_a+{\epsilon^i}_{jk}A^j_aA^k_b ,
\end{equation}
and $\Lambda^i$ is the Lagrange multiplier corresponding to $\mathcal{G}_i$ and as usual, $N^a$ is the shift vector and $\tilde{N}$ is the densitized lapse function with weight $-1$.
The diffeomorphism and Hamiltonian constraints, i.e. (\ref{diff. Cons. Ash}) and (\ref{Ham. Cons. Ash}), are the same as in the ADM formulation, i.e. (\ref{diff. Const. ADM}) and (\ref{Ham. Const. ADM}), with $q_{ab}$ and $\pi^{ab}$ expressed in terms of the Ashtekar-Barbero variables. In addition, there is an extra constraint, i.e. (\ref{Gauss Cons.}), generating the internal rotations. The canonical Hamiltonian which will be used to compute equation of motions is a linear combination of the constraints (\ref{Gauss Cons.})-(\ref{Ham. Cons. Ash}) and is expressed as
\begin{equation}\label{Canonical Ham.}
\mathbf{H}_{\text{can}}=\int_\Sigma d^3x\; (\mathcal{G}_i [\Lambda^i] + \mathcal{H}_a [N^a] + \mathcal{H} [\tilde{N}])
\end{equation}
\\
\\
We are now prepared to discuss the asymptotic behaviors of the Ashtekar-Barbero variables $(A_a^i, E^a_i)$ which satisfy requirements (i)-(iii). It is important to note that simply converting the boundary conditions imposed on the ADM-variables to the new variables does not yield a comprehensive asymptotic theory for GR written in terms of Ashtekar-Barbero variables. This is due to the existence of an additional internal $SU(2)$ frame, the asymptotic behavior of which must be determined while satisfying all consistency requirements (i)-(iii).
Reference \cite{Thiemann} provides an asymptotic analysis of the theory for $\beta= i$, which is equivalent to the R-T boundary conditions described in section \ref{Section R-T boundary conditions}. Subsequently, reference \cite{Campiglia} presents a similar analysis for a real arbitrary Barbero-Immirzi parameter $\beta$. 
In the remainder of this section, we provide a brief summary of their discussions and highlight the findings that are necessary to explain our results in the next section.

Given that the triad 1-form corresponds to the square-root of the 3-metric, it is reasonable to anticipate that the electric field will exhibit a fall-off behavior of $E^a_i= \bar{E}^a_i+\frac{\bar{f}^a_i}{r}+O(r^{-2})$. Here, $\bar{E}^a_i$ represents the densitized triad of the asymptotic 3-metric at spatial infinity. Stated differently, the associated metric of $\bar{E}^a_i$ is considered to be $\delta_{ab}$ appearing in (\ref{BC Cartesian}). It is important to note that $\bar{E}^a_i$ is not fixed in general, as it has the ability to undergo rotations in the internal space, while still maintaining $\delta_{ab}$ as its associated metric. However, ensuring a well-defined symplectic structure requires that the electric fields asymptote to a fixed densitized triad. Thus, we select the fixed, zeroth order asymptotic electric field to be $\bar{E}^a_i =\delta^a_i$, where
\[   
\delta^a_i= 
     \begin{cases}
       1 &\quad\text{if}\; (a,i)=(x,1), (y,2), (z,3),\\
       0 &\quad\text{otherwise.} \\ 
     \end{cases}
\]
Subsequently, the boundary conditions (\ref{BC Cartesian}) in terms of the Ashtekar-Barbero variables can be expressed as
\begin{equation}\label{Ashtekar BC}
    \begin{split}
        &E^a_i=\delta^a_i+\frac{\bar{f}^a_i}{r}+O(r^{-2})\\
        &A_a^i=\frac{\bar{g}^i_a}{r^2}+O(r^{-3})
    \end{split}
\end{equation}
where
and $\bar{f}^a_i$ and $\bar{g}_a^i$ are tensor fields defined on the asymptotic 2-sphere admitting the definite parity conditions \cite{Thiemann, Campiglia}
\begin{equation}\label{Ashtekar PC}
    \begin{split}
        \bar{f}^a_i\left(-\vec{\mathbf{n}}\right)=f^a_i\left(\vec{\mathbf{n}}\right), \; \; \; \bar{g}^i_a\left(-\vec{\mathbf{n}}\right)=-g^i_a\left(\vec{\mathbf{n}}\right).
    \end{split}
\end{equation}
The well-definedness of the symplectic structure (\ref{Symplectic Struc. of Ash.}) is ensured by the fall-off conditions (\ref{Ashtekar BC}) and the parity conditions (\ref{Ashtekar PC}). Specifically, the symplectic structure can be expressed as
\begin{equation}\label{div. part of symplectic str.}
\Omega(\delta_1, \delta_2) = \frac{2}{\beta} \int \frac{dr}{r} \int_{S^2} d\mathbf{\sigma} \; (\delta_1 \bar{f}^a_i \delta_2 \bar{g}^i_a - \delta_2 \bar{f}^a_i \delta_1 \bar{g}^i_a) + \text{finite}
\end{equation}
Similar to the analysis presented in section \ref{Section R-T boundary conditions}, the coefficient of the leading logarithmic singularity vanishes. This is due to the parity conditions (\ref{Ashtekar PC}) which render the term $\delta \bar{f}^a_i \delta \bar{g}^i_a$ an odd function, hence its integral over the sphere evaluates to zero.
Now it is easy to see the importance of fixing the zeroth order electric field in (\ref{Ashtekar BC}) in order to ensure the convergence of the integral (\ref{Symplectic Struc. of Ash.}). If we were to allow all possible SU(2)-rotated $\bar{E}^a_i$, the convergence of this integral could not be guaranteed. This means the requirement (i) is fulfilled.

In order to validate the requirement (iii) regarding these boundary conditions, it is necessary to obtain well-defined forms of the constraints (\ref{Gauss Cons.})-(\ref{Ham. Cons. Ash}) when the smearing functions include the Poincar\'e generators (\ref{lapse-shift Cartesian}). To achieve this, the appropriate decay behavior for $\Lambda^i$ must first be determined. Given that the leading term of $\mathcal{G}_i$ presented in (\ref{Gauss Cons.}) is an odd function with $r^{-2}$ decay, the convergence of $\mathcal{G}_i[\Lambda^i]$ is contingent upon the decay condition
\begin{equation}\label{BC for lambda}
    \begin{split}
        \Lambda^i = \frac{1}{r}\bar{\Lambda}^i+O(r^{-2})
    \end{split}
\end{equation}
where $\bar{\Lambda}^i$ are even functions defined on the asymptotic $S^2$.
It can be readily confirmed that the differentiability of $\mathcal{G}_i[\Lambda^i]$ is also ensured by (\ref{BC for lambda}).

With regards to the diffeomorphism and Hamiltonian constraints, i.e. (\ref{diff. Cons. Ash}) and (\ref{Ham. Cons. Ash}), it is observed that even after subtracting the surface destroying differentiability, the constraints only converge for translations and not for boosts and rotations. This situation necessitates modification such that: 1) the generators remain functionally differentiable and 2) the already available well-defined generator for translations remains intact up to a pure gauge.
Upon conducting a thorough and meticulous examination, it becomes evident that the issue at hand stems from the fact that, despite the fixation of the zeroth order term of the electric field as $\delta^a_i$, it continues to undergo rotation within the internal space when moving from one hypersurface to another. Put differently, $\delta^a_i$ remains unfixed during hypersurface deformations. 
For instance, it is known that under the action of the diffeomorphism constraint, the variables change according to their Lie derivative along the shift vector, i.e. $\{H_a[N^a], E^b_i\} =\mathcal{L}_{\vec{N}} E^b_i$. By examining the asymptotic behavior of this equation, it becomes evident that $\{H_a[N^a], \delta^b_i\} =\mathcal{L}_{\vec{R}} \delta^b_i$, which is not generally equal to zero. Here, $R$ represents the asymptotic rotations. This problem did not occur in the ADM variables due to the fact that $\vec{R}$ acts as a Killing vector for the asymptotic 3-metric $\delta_{ab}$, i.e. $\mathcal{L}_{\vec{R} }\delta_{ab} =0$. However, it should be noted that the fact that $\vec{R}$ is an asymptotic Killing vector does not necessarily imply that the asymptotic triad is also Lie-annihilated. It only means that the asymptotic triad is rotated within the tangent space.
Given that the Gauss constraint is primarily responsible for generating internal rotations, there is still hope of preventing these rotations by compensating for their effects through a term proportional to the Gauss constraint equipped with suitable Lagrange multipliers.
This approach has been implemented in \cite{Thiemann, Campiglia}. By subtracting a term proportional to the Gauss constraint from the diffeomorphism and Hamiltonian constraints, not only is $\delta^a_i$ fixed, but it also renders the constraints well-defined functionals. Specifically, it has been revealed that for real $\beta$, the final well-defined symmetry generators are \cite{Campiglia}
\begin{align}
        &\mathcal{H}_a[N^a]:=H_a[N^a] +s  \mathcal{G}_i[\hat{\Lambda}^i] + \text{surface terms}\label{final ex diff}\\ 
        &\mathcal{H}[\tilde{N}]:=H[\tilde{N}] - \beta \; \mathcal{G}_i [\mathring{\Lambda}^i]+ \text{surface terms}\label{final ex Ham}        
\end{align}
with suitable Lagrange multipliers
\begin{equation}
\begin{split}
& \hat{\Lambda}^i= \Lambda^i -\frac{1}{2}\epsilon_{ijk}\delta^j_a \delta^b_k b^a_b\\
& \ring{\Lambda}^i= \Lambda^i +\delta^a_i b_a.
\end{split}
\end{equation}
Although the Gauss terms have been subtracted to eliminate the source of divergence coming from boosts and rotations, still one needs some surface terms subtracted to ensure the differentiability. It should be noted that, as expected, the volume terms added to the constraints, $ s \mathcal{G}_i[\hat{\Lambda}^i]$ and $\beta \; \mathcal{G}_i [\mathring{\Lambda}^i]$ respectively, are proportional to the Gauss constraint, thereby preserving the invariance of the translation generator on the constraint surface of the Gauss constraint. 
The reader can find the surface terms in the original paper \cite{Campiglia} which are derived based on the specific boundary conditions (\ref{Ashtekar BC}) and (\ref{Ashtekar PC}). It has been verified in \cite{Thiemann, Campiglia} in more details that the standard boundary conditions (\ref{Ashtekar BC}) and (\ref{Ashtekar PC}) are preserved by hypersurfece deformations.
This analysis successfully verifies all the requirements (i)-(iii) for the standard boundary conditions. 

As the boundary conditions (\ref{Ashtekar BC}) and (\ref{Ashtekar PC}) are equivalent to R-T boundary conditions, similar to what was discussed in section \ref{Section R-T boundary conditions}, there is no scope for the supertranslation charge. 
In the following section, we present new boundary conditions within the Ashtekar-Barbero phase space, without relying on the ADM expressions, which result in non-zero supertranslation charges at spatial infinity.

\section{\textsf{New boundary conditions for Ashtekar-Barbero variables}}
In order to provide a comprehensive description of our approach for strengthening the boundary conditions, it is more practical to employ spherical coordinates $(r, x^A)$, wherein $x^A$ denotes coordinates on the sphere. In this section, we predominantly use the second set of spherical coordinates introduced in section \ref{Section H-T boundary conditions}, specifically $a,b, \cdots \in \{r, \theta , \varphi\}$,$A,B, \cdots \in \{\theta , \varphi\}$ and the antipodal map is defined as $\theta \to \pi - \theta, \varphi \to \varphi + \pi$. In these coordinates, the asymptotic conditions (\ref{Ashtekar BC}) are expressed as
\begin{equation}\label{New Ash BC}
\begin{split}
E^r_i &= r^2 \sqrt{\bar{\gamma}}\bar{\gamma}^r_i + r \sqrt{\bar{\gamma}}\bar{f}^r_i + \sqrt{\bar{\gamma}}\bar{f}^{(2)r}_i  + O(r^{-1})\\
E^A_i &= r \sqrt{\bar{\gamma}}\bar{\gamma}^A_i + \sqrt{\bar{\gamma}}\bar{f}^A_i + \sqrt{\bar{\gamma}}\bar{f}^{(2)A}_i  + O(r^{-1})\\
A^i_r &= \frac{\bar{g}^i_r}{r^2} +\frac{\bar{g}^{(2)i}_r}{r^3}+O(r^{-4})\\
A^i_B &= \frac{\bar{g}^i_B}{r}+ \frac{\bar{g}^{(2)i}_B}{r^2}+O(r^{-3})
\end{split}
\end{equation}
where $\bar{\gamma}$ is the determinant of the unit metric $\bar{\gamma}_{AB}$ on the sphere, $\bar{\gamma}^a_i$ are asymptotic triads satisfying $\bar{\gamma}^a_i \bar{\gamma}^b_i = \bar{\gamma}^{ab}=\text{diag}(1, \bar{\gamma}^{AB})$.
The behavior of the inverse of $E^a_j$ in the asymptotic region is given by
\begin{equation}\label{Asym. Exp. Inverse E}
\begin{split}
E_r^i &= \frac{1}{r^2} \frac{\bar{\gamma}_r^i}{\sqrt{\bar{\gamma}}} + \frac{1}{r^3} \frac{\bar{f}_r^i}{\sqrt{\bar{\gamma}}} + O(r^{-4})\\
E_A^i &= \frac{1}{r} \frac{\bar{\gamma}_A^i}{\sqrt{\bar{\gamma}}} + \frac{1}{r^2} \frac{\bar{f}_A^i}{\sqrt{\bar{\gamma}}} + O(r^{-3})
\end{split}
\end{equation}
Here, $\bar{\gamma}_a^i$ represents the inverse of  $\bar{\gamma}^a_i$ satisfying $\bar{\gamma}_a^i \bar{\gamma}_b^i = \bar{\gamma}_{ab}=\text{diag}(1, \bar{\gamma}_{AB})$. 
It should be noted that when deriving the first two equations in (\ref{New Ash BC}) and (\ref{Asym. Exp. Inverse E}), one must consider that the electric field and its inverse follow the coordinate transformation rules of tensor densities, rather than those of tensor fields. 
Furthermore, from the requirement $E^i_a E_i^b=\delta^b_a$ and the relations $\bar{\gamma}^a_i \bar{\gamma}_b^i=\delta^a_b$ and $\bar{\gamma}^a_i \bar{\gamma}_a^j=\delta^j_i$,  we can determine $\bar{f}_b^i$ in terms of $\bar{f}^a_i$ as
\begin{equation}\label{inverse of little f}
\bar{f}_b^i=- \bar{\gamma}_b^j \bar{\gamma}^i_a \bar{f}^a_j
\end{equation}
Two properties of $\bar{\gamma}^a_i$ that will be significant for subsequent calculations are 
\begin{align}\label{properties of gammas}
\bar{D}_A \bar{\gamma}^i_r=&\bar{\gamma}^i_A\\
\bar{D}_A \bar{\gamma}^i_B =& -\bar{\gamma}_{AB} \bar{\gamma}^i_r 
\end{align}
Under the action of the antipodal map $\bar{\gamma}_a^i$ are
\begin{equation}\label{parity of gamma}
\bar{\gamma}_i^r \sim \bar{\gamma}_i^\varphi = \text{odd}\;\;\;\; \bar{\gamma}_i^\theta = \text{even}
\end{equation}
and their lower spacetime index have the same parity because
\begin{equation}\label{inverse of gamma}
\bar{\gamma}^i_r=\bar{\gamma}^r_i,\;\; \bar{\gamma}^i_\theta=\bar{\gamma}^\theta_i,\;\; \bar{\gamma}^i_\varphi = \bar{\gamma} \bar{\gamma}^\varphi_i
\end{equation}

Moreover, since $\{\bar{\gamma}^i_r, \bar{\gamma}^i_\theta, \bar{\gamma}^i_\varphi \}$ forms a basis for the internal space, it proves convenient to use the expression $\bar{f}^a_i= \bar{F}^a_b \bar{\gamma}^b_i$, $\bar{g}_b^i= \bar{G}^a_b \bar{\gamma}_a^i$ because in the end we will associate desired parities to the components
\begin{align}
&\bar{F}^a_b := \bar{f}^a_i \bar{\gamma}^i_b \label{Components F}\\
&\bar{G}^a_b := \bar{g}^i_b \bar{\gamma}_i^a \label{Components G}
\end{align} 
In general, for any field $\bar{v}^a_i$ (or $\bar{v}_a^i$) on the asymptotic 2-sphere, its components in this basis are defined as $\bar{v}^a_b := \bar{v}^a_i \bar{\gamma}^i_b$ (by $\bar{v}^b_a := \bar{v}_a^i \bar{\gamma}_i^b$).
 
 The asymptotic behavior for lapse and shift are still assumed to be (\ref{lapse Spherical}) and (\ref{shift Spherical}). Since the smearing function for the Hamiltonian constraint (\ref{Ham. Cons. Ash}) is considered to be a scalar density of weight $-1$, we need to derive the asymptotic behavior of $\tilde{N}$ based on that of $N$ and the relation $\tilde{N}=\frac{N}{\sqrt{q}}$, namely,
 \begin{equation}\label{lapse density}
\tilde{N} = \frac{1}{r} \frac{b}{\sqrt{\bar{\gamma}}} + \frac{1}{r^2} \frac{\tilde{f}}{\sqrt{\bar{\gamma}}} + O(r^{-3})
\end{equation}
The relation between $f$ in (\ref{lapse Spherical}) and $\tilde{f}$ in (\ref{lapse density}) is given by
\begin{equation}
\tilde{f} = f - \frac{b}{2} (\bar{F}^r_r+ \bar{F}^A_A)
\end{equation}
where we have used the asymptotic expansion of the determinant of the metric
\begin{equation}\label{expansion of det q}
q = \det E^a_i = r^4 \bar{\gamma} + r^3 \bar{\gamma} (\bar{F}^r_r+ \bar{F}^A_A) + O(r^2)
\end{equation}
From this point forward, the smearing functions we use for the diffeomorphism and Hamiltonian constraints are (\ref{shift Spherical}) and (\ref{lapse density}), respectively.
 
Moving forward, we will assume that we are always working in a coordinate system in which the equation $\bar{\gamma}_{AB} \bar{F}^A_r + \bar{F}^r_B = 0$ holds. This assumption is the same as the one stated in section \ref{Section H-T boundary conditions}, where we set $\bar{h}_{rA}=0$ \cite{Henneaux}. In fact, this assumption for mixed radial-angular components of metric was imposed also in \cite{Compere:2011ve}. The only cost we must pay to ensure that we remain in this coordinate when transitioning from one hypersuface to the next is to fix $I^A$ in the shift vector (\ref{shift Spherical}) (for more details, refer to equations (\ref{Var. assumption 1}) and (\ref{Fixed I})).
 
For future reference, it is necessary to obtain the asymptotic expansion of the spin connection and extrinsic curvature, which are parts of the definition of the Ashtekar-Barbero connection (\ref{Ashtekar connection}), as well as the curvature (\ref{Fab}) that appears in the constraints (\ref{diff. Cons. Ash}) and (\ref{Ham. Cons. Ash}). These quantities can be expanded in spherical coordinates as
\begin{align}
&K^i_r = \frac{\bar{k}^i_r}{r^2}+O(r^{-3}), \;\;\;\;\;\; K^i_B = \frac{\bar{k}^i_B}{r}+O(r^{-2})\\
&\Gamma^i_r = \frac{\bar{\Gamma}^i_r}{r^2}+O(r^{-3}),\;\;\;\;\; \;\; \Gamma^i_B = \frac{\bar{k}^i_B}{r}+O(r^{-2})\\
F_{r A}= \frac{1}{r^2}& \left(-\bar{g}^i_A  - \partial_A \bar{g}^i_r \right) + O(r^{-3}),\;\;\;\;  F_{A B}= \frac{1}{r} \left(\partial_A \bar{g}^i_B - \partial_B \bar{g}^i_A \right) + O(r^{-2})
\end{align}
By using equations (\ref{Spin connection}) and (\ref{New Ash BC}) and and performing a lengthy calculation, we obtain the leading order term of the spin connection as 
\begin{align}\label{LOT radial spin connection}
\bar{\Gamma}^i_r = &\frac{1}{2 \sqrt{\bar{\gamma}}}  \left[2 \epsilon^{CB} \bar{\gamma}_{AB} \bar{F}^A_r -\epsilon^{CA}\bar{D}_A (\bar{F}^r_r - \bar{F}^B_B) \right] \bar{\gamma}^i_C\nonumber\\
&+\frac{1}{2 \sqrt{\bar{\gamma}}} \epsilon^{BA} \left[\bar{D}_A (\bar{\gamma}_{CB} \bar{F}^C_r + \bar{F}^r_B) + \bar{\gamma}_{CB}\bar{F}^C_A \right] \bar{\gamma}^i_r\nonumber\\
=& \; \bar{\Gamma}^C_r \bar{\gamma}^i_C + \bar{\Gamma}^r_r \bar{\gamma}^i_r
\end{align}
\begin{align}\label{LOT angular spin connection}
\bar{\Gamma}^i_B = & \frac{1}{2 \sqrt{\bar{\gamma}}} \left[\epsilon^{DA}\bar{\gamma}_{AB}(\bar{F}^C_C - \bar{F}^r_r) - 2 \epsilon^{DC} \bar{\gamma}_{AC} (\bar{D}_B \bar{F}^A_r) \right] \bar{\gamma}^i_D \nonumber\\
& 
+\frac{1}{2 \sqrt{\bar{\gamma}}} \left[2\epsilon^{DA}\bar{\gamma}_{BD} \bar{\gamma}_{AC} \bar{F}^C_r - \epsilon^{DA} \bar{\gamma}_{BD} \bar{D}_A (\bar{F}^r_r + \bar{F}^C_C) + 2 \epsilon^{DA}\bar{\gamma}_{CD}\bar{D}_A \bar{F}^C_B \right] \bar{\gamma}^i_r \nonumber\\
=& \; \bar{\Gamma}^D_B \bar{\gamma}^i_D + \bar{\Gamma}^r_B \bar{\gamma}^i_r
\end{align}
Along with equation (\ref{Ashtekar connection}), these results allow us to determine the leading terms of $K^i_a$ as
\begin{equation}\label{LOT extrinsic curvature}
\bar{k}^i_a= \beta^{-1}(\bar{g}^i_a- \bar{\Gamma}^i_a) = \beta^{-1}(\bar{G}^b_a \bar{\gamma}_b^i - \bar{\Gamma}^i_a) 
\end{equation}
Finally, by replacing the expressions (\ref{LOT radial spin connection}) and (\ref{LOT angular spin connection}) into (\ref{LOT extrinsic curvature}), we can obtain the explicit formula for $\bar{k}^i_a$ in terms of $\bar{F}^a_b$ and $\bar{G}^a_b$.

\subsection{\textsf{Explicit form}}\label{Section Explicit form}
We are now prepared to present the new boundary conditions for the Ashtekar-Barbero variables in the form of a theorem. The subsequent sections of this paper will be dedicated to the thorough analysis of the implications and substance of this theorem.
\begin{theorem*}
The following boundary conditions meet both requirements (i) and (ii). They also fulfill requirement (iii) for spacetime translations and result in non-zero supertranslation charges.
The boundary conditions consist of the decay conditions (\ref{New Ash BC}) and the parity conditions
\begin{align}
&\bar{F}^\theta_r \sim \bar{F}^\theta_\varphi \sim \bar{F}^\varphi_\theta = \text{even} \label{PC 1}\\
&\bar{F}^\varphi_r \sim \bar{F}^\theta_\theta \sim \bar{F}^\varphi_\varphi \sim \bar{G}^r_r = \text{odd} \label{PC 2}\\
&(\bar{F}^r_r - \bar{F}^A_A) \sim  (\bar{G}^\theta_\theta + \bar{G}^r_r) \sim  (\bar{G}^\varphi_\varphi + \bar{G}^r_r)= \text{even} \label{PC 3}\\
&\bar{G}^\theta_r + \frac{1}{2 \sqrt{\bar{\gamma}}} \bar{D}_\varphi (\bar{F}^r_r - \bar{F}^A_A) = \text{odd} \label{PC 4}\\
&\bar{G}_\theta^r + \frac{1}{2 \sqrt{\bar{\gamma}}} \bar{D}_\varphi (\bar{F}^r_r - \bar{F}^A_A) = \text{odd} \label{PC 5}\\
&\bar{G}^\varphi_r - \frac{1}{2 \sqrt{\bar{\gamma}}} \bar{D}_\theta (\bar{F}^r_r - \bar{F}^A_A) = \text{even} \label{PC 6}\\
&\bar{G}_\varphi^r - \frac{\sqrt{\bar{\gamma}}}{2 } \bar{D}_\theta (\bar{F}^r_r - \bar{F}^A_A)  = \text{even}\label{PC 7}\\
&\bar{G}^\theta_\varphi + \frac{\sqrt{\bar{\gamma}}}{2} \left( \bar{F}^r_r - \bar{F}^A_A \right) = \text{odd}\label{PC 8}\\
&\bar{G}_\theta^\varphi - \frac{1}{2 \sqrt{\bar{\gamma}}} \left(\bar{F}^r_r  -\bar{F}^A_A \right) = \text{odd} \label{PC 9}
\end{align}
Moreover, we have assumed
\begin{align}
&\bar{\gamma}_{AB} \bar{F}^A_r + \bar{F}^r_B = 0 \label{assumption 1}\\
&\bar{\gamma}_{B[E} \delta \bar{F}^B_{D]} = 0\label{assumption 2}
\end{align}
The restrictions we have in our Lagrange multipliers are
\begin{align}
&W = \text{odd}, \;\;\;\; \tilde{f} = \text{even} \label{parity of W and tilde f}\\
&I_D = \partial_D W -\beta \; b \left[ \frac{s}{\beta}(\bar{k}^r_D + \bar{\gamma}_{BD} \bar{k}^B_r) - \frac{\epsilon^{AC} }{\sqrt{\bar{\gamma}}}  \bar{\gamma}_{AD} \bar{\gamma}_{BC} \bar{F}^B_r  \right]\label{I fixed}\\
&2 \bar{\Lambda}^r  = -s \; \frac{\epsilon^{DA} }{\sqrt{\bar{\gamma}} } \bar{D}_A \left( b\;   [\bar{k}^r_D + \bar{\gamma}_{BD} \bar{k}^B_r]  \right)- \beta \left[ \bar{F}^A_r (\bar{D}_A b) - 2 b \bar{D}_A \bar{F}^A_r + b  \; \frac{\epsilon^{AB}}{\sqrt{\bar{\gamma}}}  \bar{\gamma}_{BC} \bar{G}^C_A  \right] \label{Lambdar}\\ 
&\bar{\Lambda}_\theta  + \beta \frac{b}{2} \bar{D}_\theta (\bar{F}^r_r - \bar{F}^A_A) = \text{odd}\label{Lambda theta}\\
& \bar{\Lambda}_\varphi  + \beta \frac{b}{2} \bar{D}_\varphi (\bar{F}^r_r - \bar{F}^A_A) = \text{even}\label{Lambda phi}
\end{align}
where $\bar{\Lambda}^a:= \bar{\Lambda}^i \bar{\gamma}_i^a$.
\end{theorem*}
 Note that obviously the new boundary conditions differ from the standard ones. Specifically, in terms of $\bar{F}^a_b, \bar{G}^a_b$ the standard parity conditions \ref{Ashtekar PC} is stated as
\begin{equation}
\begin{split}
&\bar{F}^a_r = \bar{f}^a_i \bar{\gamma}^i_r = \text{odd}, \;\;\; \bar{F}^a_\theta = \bar{f}^a_i \bar{\gamma}^i_\theta = \text{even}, \;\;\; \bar{F}^a_\varphi = \bar{f}^a_i \bar{\gamma}^i_\varphi = \text{odd},\\
&\bar{G}^r_a = \bar{g}_a^i \bar{\gamma}_i^r = \text{even}, \;\;\; \bar{G}^\theta_a = \bar{g}_a^i \bar{\gamma}_i^\theta = \text{odd}, \;\;\; \bar{G}^\varphi_a = \bar{g}_a^i \bar{\gamma}_i^\varphi = \text{even}
\end{split}
\end{equation}
where all the canonical variables have definite parities. Here to read the parities we have the relations (\ref{parity of gamma}) and (\ref{inverse of gamma}) in mind.  
Furthermore, the parity condition $\bar{\Lambda}^i = \text{even}$ in the standard boundary conditions is equivalent to $\bar{\Lambda}^r \sim \bar{\Lambda}^\varphi =\text{even}$ and $\bar{\Lambda}^\theta =\text{odd}$, while the equations (\ref{Lambdar})-(\ref{Lambda phi}) tells us that all $\bar{\Lambda}^a$ do have both even and odd pieces in the new setting of boundary conditions. Moreover, in the standard parity conditions $I^A$ is not fixed, while here it is through the equation (\ref{I fixed}).

The rest of this section is dedicated to prove this theorem. First, we show that these boundary conditions gives us a well-defined symplectic structure (\ref{Symplectic Struc. of Ash.}). As we already saw in (\ref{div. part of symplectic str.}), since the asymptotic triads are supposed to be fixed, the divergent part of the symplectic structure is
\begin{equation}
\begin{split}
& \frac{2}{\beta} \int \frac{dr}{r} \int_{S^2} d\mathbf{\sigma} \; (\delta_1 \bar{f}^a_i \delta_2 \bar{g}^i_a - \delta_2 \bar{f}^a_i \delta_1 \bar{g}^i_a)\\
= & \frac{2}{\beta} \int \frac{dr}{r} \int_{S^2} d\mathbf{\sigma} \; (\delta_1 \bar{F}^a_b \delta_2 \bar{G}^b_a - \delta_2 \bar{F}^a_b \delta_1 \bar{G}^b_a) \\
= & \frac{2}{\beta} \int \frac{dr}{r} \int_{S^2} d\mathbf{\sigma} \; \left(\delta_1 \bar{F}^r_r \delta_2 \bar{G}^r_r + \delta_1 \bar{F}^r_A \delta_2 \bar{G}^A_r + \delta_1 \bar{F}^A_r \delta_2 \bar{G}^r_A + \delta_1 \bar{F}^A_B \delta_2 \bar{G}^B_A  - (\delta_1 \leftrightarrow \delta_2)\right)\\
= & \frac{2}{\beta} \int \frac{dr}{r} \int_{S^2} d\mathbf{\sigma} \; \left(\delta_1 (\bar{F}^r_r - \bar{F}^A_A) \delta_2 \bar{G}^r_r + \delta_1 \bar{F}^A_A \delta_2 \bar{G}^r_r + \delta_1 \bar{F}^A_r \delta_2 (\bar{G}^r_A-\bar{\gamma}_{AB} \bar{G}^B_r)  + \delta_1 \bar{F}^A_B \delta_2 (\bar{G}^B_A)_{\text{sym}}   - (\delta_1 \leftrightarrow \delta_2)\right)\\
= & \frac{2}{\beta} \int \frac{dr}{r} \int_{S^2} d\mathbf{\sigma} \; \left(\delta_1 \bar{F}^A_A \delta_2 \bar{G}^r_r  + \delta_1 \bar{F}^A_B \delta_2 [(\bar{G}^B_A)_{TL} +\frac{1}{2}\bar{G}^C_C \delta^B_A ]   - (\delta_1 \leftrightarrow \delta_2)\right)\\
= & \frac{2}{\beta} \int \frac{dr}{r} \int_{S^2} d\mathbf{\sigma} \; \left(\delta_1 \bar{F}^A_A \delta_2 (\bar{G}^r_r +\frac{1}{2}\bar{G}^C_C ) + \delta_1 (\bar{F}^A_B)_{TL} \delta_2 (\bar{G}^B_A)_{TL}  - (\delta_1 \leftrightarrow \delta_2)\right)  \\
= & \; 0
\end{split}
\end{equation}
Here, in the first step we have used (\ref{Components F}), (\ref{Components G}) and $\bar{\gamma}^a_i \bar{\gamma}_b^i=\delta^a_b$. In the second step we have separated the radial and angular components. In the third step we have added and subtracted $\delta_1 \bar{F}^A_A \delta_2 \bar{G}^r_r$, the we used (\ref{assumption 1}) to conclude that $\delta_1 \bar{F}^r_A \delta_2 \bar{G}^A_r + \delta_1 \bar{F}^A_r \delta_2 \bar{G}^r_A= \delta_1 \bar{F}^A_r \delta_2 (\bar{G}^r_A-\bar{\gamma}_{AB} \bar{G}^B_r) $ and we also used (\ref{assumption 2}) to see that only the symmetric part of $\bar{G}^B_A$ contributes in it as $\delta_1 \bar{F}^A_B \delta_2 \bar{G}^B_A = \delta_1 \bar{F}^A_B \delta_2 (\bar{G}^B_A)_{\text{sym}}$ where by $(\bar{G}^B_A)_{\text{sym}}$ we mean the symmetric part of the 2 by 2 matrix $\bar{G}^B_A$. In the fourth step, first we used the parity conditions (\ref{PC 1}) to see that $\delta_1 (\bar{F}^r_r - \bar{F}^A_A) \delta_2 \bar{G}^r_r = \text{odd}$ and also the parity conditions (\ref{PC 4})-(\ref{PC 7}) to conclude that  $\delta_1 \bar{F}^A_r \delta_2 (\bar{G}^r_A-\bar{\gamma}_{AB} \bar{G}^B_r) = \text{odd}$. These two odd terms vanish because integration of an odd function over 2-sphere is equal to zero. Then we have also split the symmetric matrix $(\bar{G}^B_A)_{\text{sym}}$ into the trace piece and traceless piece $(\bar{G}^B_A)_{TL}$, i.e. $(\bar{G}^B_A)_{\text{sym}}= (\bar{G}^B_A)_{TL} +\frac{1}{2}\bar{G}^C_C \delta^B_A$. In the fifth step we have factored $\bar{F}^A_A$  out of the terms including it and noted that only the traceless part of the symmetric matrix $\bar{F}^A_B$ contributes in the term $\delta_1 \bar{F}^A_B \delta_2 (\bar{G}^B_A)_{TL}$. Finally, by using the parity conditions (\ref{PC 1})-(\ref{PC 3}), (\ref{PC 8}) and (\ref{PC 9}) we see that the two remaining terms are both odd functions and so their integrations vanish. 

Therefore, the parity conditions introduced in the theorem ensure convergence of the symplectic structure and the requirement (i) is satisfied.

\subsection{\textsf{Constraints}}\label{Section Constraints}
The newly established parity conditions do ensure the finite nature of the symplectic form, but they alone do not guarantee the cancellation of the divergent components in the boost charges and angular momentum, in contrast to the parity conditions presented in \cite{RT}. Therefore, these conditions must be complemented by additional asymptotic restrictions in order to achieve the finiteness of the charges.

As explained in section \ref{Section H-T boundary conditions}, the approach employed in \cite{Henneaux} to solve this issue involves setting the leading terms of the constraints to zero. These supplementary conditions are relatively moderate. In this study, while working with Ashtekar-Barbero variables, our objective is to investigate whether this strategy leads to the elimination of divergence in surface terms. For this purpose, in the current section, we derive the leading terms of the constraints in terms of the Ashtekar-Barbero variables and set them to zero. Additionally, we analyze the behavior of the constraints and their variations at infinity.

With the given boundary conditions (\ref{New Ash BC}) and in spherical coordinates, the constraints (\ref{Gauss Cons.})-(\ref{Ham. Cons. Ash}) exhibit the following decay behavior
\begin{align}
\mathcal{G}_i =& \; \bar{\mathcal{G}}_i + O(r^{-1})\\
H_r = \frac{1}{r}\bar{H}_r + O(& r^{-2}),\;\;\; H_A = \bar{H}_A + O(r^{-1})\\
H =& \;r \bar{H} + O(1) 
\end{align}
The strengthened boundary conditions require the absence of leading divergences in the constraints, thus imposing the conditions:
\begin{equation}\label{mild conditions}
\bar{\mathcal{G}}_i = \bar{H}_r = \bar{H}_A = \bar{H} = 0
\end{equation}
As in this section we wish to thoroughly discuss aspects related to the constraints, we will derive their actions on the canonical variables and also examine the surface terms that affect their differentiability, which will be addressed in section \ref{Section Asymptotic charges} when the asymptotic charges are discussed. 

Before delving into the details, it is important to recall from section \ref{Section Standard BC Ash.} that to ensure the finiteness of the volume part of the diffeomorphism and Hamiltonian constraints, a term proportional to the Gauss constraint needs to be added to their expressions (see equation (\ref{final ex diff}) and (\ref{final ex Ham}) respectively). In spherical coordinates, the Lagrange multipliers $\hat{\Lambda}^i$ and $\mathring{\Lambda}^i$ are given by:
\begin{align}
&\hat{\Lambda}^i =  \Lambda^i + \frac{1}{2} \frac{\epsilon^{BC}}{\sqrt{\bar{\gamma}}} \bar{\gamma}_{AC} \left(-2 Y^A \bar{\gamma}^j_B +  (\bar{D}_BY^A) \bar{\gamma}^j_r \right)\label{Lagrange Multi. diff}\\
&\mathring{\Lambda}^i = \Lambda^i + (b \bar{\gamma}^r_j + (\partial_A b) \bar{\gamma}^A_j).\label{Lagrange Multi. Ham}
\end{align}
In the subsequent discussion, we examine the behavior of each constraint individually.

\subsubsection{\textsf{Gauss constraint}}
After some calculation, one can obtain the explicit expression of the conditions $\bar{\mathcal{G}}_i = 0$ in (\ref{mild conditions}) as equivalent to the following equations
\begin{align}
&\bar{\mathcal{G}}^r = \frac{2}{\beta} \left(\sqrt{\bar{\gamma}} (\bar{F}^r_r - \bar{F}^A_A) + \sqrt{\bar{\gamma}}(\bar{D}_A \bar{F}^A_r) - \epsilon^{AB} \bar{G}^C_A \bar{\gamma}_{BC} \right)=0 \label{LOT radial Gauss}\\
&\bar{\mathcal{G}}^A = \frac{2}{\beta} \left(\sqrt{\bar{\gamma}} (\bar{F}^r_B \bar{\gamma}^{AB} + \bar{F}^A_r) + \sqrt{\bar{\gamma}} \bar{\gamma}^{AB} (\bar{D}_C \bar{F}^C_B) + \epsilon^{AB} (\bar{G}^C_r \bar{\gamma}_{BC}-\bar{G}^r_B) \right)\nonumber \\
&\;\;\;\;\; = \frac{2}{\beta} \left( \sqrt{\bar{\gamma}} \bar{\gamma}^{AB} (\bar{D}_C \bar{F}^C_B) + \epsilon^{AB} (\bar{G}^C_r \bar{\gamma}_{BC}-\bar{G}^r_B) \right) \label{LOT angular Gauss}
\end{align} 
where $\bar{\mathcal{G}}_i = \bar{\mathcal{G}}^r \bar{\gamma}^i_r + \bar{\mathcal{G}}^A \bar{\gamma}^i_A$, and to obtain the second line of (\ref{LOT angular Gauss}), we make use of the assumption (\ref{assumption 1}). It is important to ensure the convergence of the smeared constraint.
In the standard boundary conditions discussed in section \ref{Section Standard BC Ash.}, it is required to assign definite even parity to $\bar{\Lambda}^i$ in order to ensure the finiteness of (\ref{Gauss Cons.}). However, in this case, we do not need to restrict the parity of $\bar{\Lambda}^i$ as the divergent part of (\ref{Gauss Cons.}) has already been removed, as represented by the equation
\begin{equation}
\mathcal{G}_i [\Lambda^i] = \int_\Sigma dr d\theta d\varphi \; \mathcal{G}_i \Lambda^i =
 \int \frac{dr}{r} \int_{S^2} d\mathbf{\sigma} \; (\bar{\Lambda}^i \bar{\mathcal{G}}_i) + \text{finite} = \text{finite}
\end{equation}
where we have used the condition $\bar{\mathcal{G}}_i = 0$.

Now, it is necessary to verify the differentiability of the constraint. The variation of the smeared Gauss constraint can be represented as 
\begin{equation}
\begin{split}
\delta \mathcal{G}_i [\Lambda^i] =& \frac{2}{\beta} \int_\Sigma d^3x\; \Lambda^i \left(\partial_a \delta E^a_i + \epsilon_{ijk} (\delta A^j_a) E^a_k + \epsilon_{ijk} A^j_a (\delta E^a_k) \right)\\
=&
\frac{2}{\beta} \int_\Sigma d^3x\; \left(- \partial_a \Lambda^k + \epsilon_{ijk}  A^j_a \Lambda^i \right) (\delta E^a_k) + \left(\epsilon_{ijk} \Lambda^i E^a_k \right) (\delta A_a^j) + \frac{2}{\beta} \oint dS_a (\Lambda^i \delta E^a_i)
\end{split}
\end{equation}
By expressing the above variation in the form
\begin{equation}
\delta \mathcal{G}_i [\Lambda^i] = \Omega(\delta_{\Lambda}, \delta ) + \mathcal{B}_{\Lambda} (\delta A^i_a, \delta E^a_i)= \frac{2}{\beta} \int_\Sigma d^3x\; \left((\delta E^a_i)(\delta_{\Lambda} A^i_a) - (\delta A^i_a) (\delta_{\Lambda} E^a_i) \right)+ \mathcal{B}_{\Lambda} (\delta A^i_a, \delta E^a_i)
\end{equation}
we can determine the variation of the canonical variables under the action of the Gauss constraint, as well as the surface term that needs to be subtracted in order to obtain a functionally differentiable generator for internal rotations. Specifically, we find
\begin{align}
&\delta_{\Lambda} A^i_a = - \partial_a \Lambda^i + \epsilon_{ijk} \Lambda^j  A^k_a \label{Var. of A under Guass Const.}\\
&\delta_{\Lambda} E^a_i =  \epsilon_{ijk} \Lambda^j E^a_k \label{Var. of E under Guass Const.}\\
&\mathcal{B}_{\Lambda} (\delta A^i_a, \delta E^a_i) = \frac{2}{\beta} \oint dS_a (\Lambda^i \delta E^a_i) \label{BT of Guass. const.}
\end{align}
We will address the issue arising from the surface term in section \ref{Section Asymptotic charges}.

\subsubsection{\textsf{Diffeomorphism constraint}}
The conditions $\bar{H}_r = \bar{H}_A = 0$ stated in (\ref{mild conditions}) can be expressed explicitly as
\begin{align}
&\bar{H}_r = \frac{2 s}{\beta} \sqrt{\bar{\gamma}}\left(\bar{G}^A_A + 2 \bar{G}^r_r + \bar{D}_A \bar{G}^A_r \right) =0 \label{LOT radial Diff}\\
&\bar{H}_A = -\frac{2 s}{\beta} \sqrt{\bar{\gamma}}\left(\bar{D}_A \bar{G}^r_r - \bar{G}^B_r \bar{\gamma}_{AB} + \bar{D}_A \bar{G}^B_B - \bar{D}_B \bar{G}^B_A \right) =0 \label{LOT angular Diff}
\end{align}
In order to check the convergence of the smeared constraint (\ref{diff. Cons. Ash}), we calculate its asymptotic expansion
\begin{align}\label{div. diff.}
H_a [N^a] &= \int_\Sigma dr d\theta d\varphi \; \left(H_r N^r + H_A N^A \right) \nonumber\\
&= \int \frac{dr}{r} \int_{S^2} d\mathbf{\sigma} \; (W \bar{H}_r) + \int  dr \int_{S^2} d\sigma (Y^A \bar{H}_A + O(r^{-1})) + \text{finite}
\end{align}
It is evident that even with the imposition of the conditions $\bar{H}_r = \bar{H}_A = 0$, the constraint $H_a[N^a]$ diverges due to the logarithmic divergence arising from the second integral in (\ref{div. diff.}). This issue was already encountered in section \ref{Section Standard BC Ash.}, where applying the standard parity conditions allowed the removal of the divergent parts including $W \bar{H}_r$ and $Y^A \bar{H}_A$, but the logarithmic divergence in the second integral of (\ref{div. diff.}) persisted. As mentioned previously, this problem can be resolved without relying on any specific parity condition by working with $\mathcal{H}_a[N^a]$ defined in (\ref{final ex diff}) alongside the Lagrange multiplier (\ref{Lagrange Multi. diff}) instead of $H_a[N^a]$. For a more detailed explanation, the interested reader can refer to \cite{Thiemann, Campiglia}.

Although $\mathcal{H}_a[N^a]$ provides us with a finite generator for the symmetries, it is still necessary to ensure its differentiability. Let us compute the variation of $\mathcal{H}_a[N^a]$ to obtain the surface terms mentioned in (\ref{final ex diff}), as well as the variations of the canonical variables under its action

\begin{equation}
\begin{split}
\delta (\mathcal{H}_a[N^a]) =& \frac{-2 s}{\beta} \delta \int_\Sigma d^3x\; N^a \left(F^i_{ab} E^b_i - A^i_a \mathcal{G}_i \right) + s \delta \mathcal{G}_i[\hat{\Lambda}^i]\\
=&
 \frac{-2 s}{\beta} \delta \int_\Sigma d^3x\; N^a \left([\partial_a A^j_b - \partial_b A^j_a] E^b_i - A^j_a \partial_b E^b_j \right) +s \delta \mathcal{G}_i[\hat{\Lambda}^i]\\
=&
\frac{-2 s}{\beta} \int_\Sigma d^3x\; N^a \left(2 (\partial_{[a}  \delta A^j_{b]})  E^b_i + 2 (\partial_{[a} A^j_{b]}) \delta E^b_i - (\delta A^j_a) \partial_b E^b_j - A^j_a \partial_b \delta E^b_j \right) + s \delta \mathcal{G}_i[\hat{\Lambda}^i]\\
=& 
\frac{-2 s}{\beta} \int_\Sigma d^3x\; \left((\delta  E^a_i)(\mathcal{L}_{\vec{N}} A^i_a) - (\delta  A_a^i)(\mathcal{L}_{\vec{N}} E^a_i) \right) \\
&- 
\frac{2 s}{\beta} \int_{S^2} dS_a\; \left(N^a E^b_i \delta A^i_b - N^b E^a_i \delta A^i_b - N^b A^i_b \delta  E^a_i \right) +s \delta \mathcal{G}_i[\hat{\Lambda}^i]\\
=& 
\frac{-2 s}{\beta} \int_\Sigma d^3x\; \left((\delta  E^a_i)(\mathcal{L}_{\vec{N}} A^i_a - \epsilon_{ijk} \hat{\Lambda}^j A^k_a) - (\delta  A_a^i)(\mathcal{L}_{\vec{N}} E^a_i - \epsilon_{ijk} \hat{\Lambda}^j E^a_k) \right) \\
&- 
\frac{2 s}{\beta} \oint_{S^2} dS_a\; \left(N^a E^b_i \delta A^i_b - N^b E^a_i \delta A^i_b - N^b A^i_b \delta  E^a_i - \hat{\Lambda}^i \delta E^a_i \right)
\end{split}
\end{equation}
 By expressing the above variation in the form
\begin{equation}
\delta \mathcal{H}_a [N^a] = \Omega(\delta_{\vec{N}}, \delta ) + \mathcal{B}_{\vec{N}} (\delta A^i_a, \delta E^a_i)= \frac{2}{\beta} \int_\Sigma d^3x\; \left((\delta E^a_i)(\delta_{\vec{N}} A^i_a) - (\delta A^i_a) (\delta_{\vec{N}} E^a_i) \right)+ \mathcal{B}_{\vec{N}} (\delta A^i_a, \delta E^a_i)
\end{equation}
We can determine the variation of the canonical variables under the action of the diffeomorphism constraint, as well as the surface term that must be subtracted to obtain a functionally differentiable generator for spatial diffeomorphisms. Specifically
\begin{align}
&\delta_{\vec{N}} A^i_a = -s (\mathcal{L}_{\vec{N}} A^i_a - \epsilon_{ijk} \hat{\Lambda}^j A^k_a) \label{Var. of A under diff Const.}\\
&\delta_{\vec{N}} E^a_i =  -s (\mathcal{L}_{\vec{N}} E^a_i - \epsilon_{ijk} \hat{\Lambda}^j E^a_k) \label{Var. of E under diff Const.} \\
&\mathcal{B}_{\vec{N}} (\delta A^i_a, \delta E^a_i) = - \frac{2 s}{\beta} \oint_{S^2} dS_a\; \left(N^a E^b_i \delta A^i_b - N^b E^a_i \delta A^i_b - N^b A^i_b \delta  E^a_i - \hat{\Lambda}^i \delta E^a_i \right)\label{BT of diff. const.}
\end{align}
we will treat the issue coming from the surface term in section \ref{Section Asymptotic charges}.

\subsubsection{\textsf{Hamiltonian constraint}}
The explicit expression of the condition $\bar{H} = 0$ in (\ref{mild conditions}) is
\begin{equation}
\bar{H} = -2 \sqrt{\bar{\gamma}} \bar{D}_A \left(\epsilon^{AB} (\bar{G}^C_r \bar{\gamma}_{BC}- \bar{G}^r_B) \right) =0 \label{LOT Ham.}\\
\end{equation}
To check convergence of the smeared constraint (\ref{Ham. Cons. Ash}), we compute the asymptotic expansion of it, namely,
\begin{align}\label{div. Ham.}
H [\tilde{N}] &= \int_\Sigma dr d\theta d\varphi \; \tilde{N} H  
=  \int  dr \int_{S^2} d\sigma (b \bar{H} + O(r^{-1})) + \text{finite}
\end{align}
Despite implementing the conditions $\bar{H}=0$, it is apparent that the constraint $H [\tilde{N}]$ becomes divergent due to the logarithmic divergence arising from the $O(r^{-1})$ term in equation (\ref{div. Ham.}). It is noteworthy to mention that a similar problem was encountered in section \ref{Section Standard BC Ash.}, where the utilization of standard parity conditions allowed for the elimination of divergent parts, including $b \bar{H}$, while the logarithmic divergence in the second integral of equation (\ref{div. Ham.}) persisted. As previously stated, this issue can be resolved without relying on any specific parity condition by working with $\mathcal{H}[N]$ as defined in  (\ref{final ex Ham}) and incorporating the Lagrange multiplier from  (\ref{Lagrange Multi. Ham}). For a more comprehensive explanation, we recommend referring to \cite{Thiemann, Campiglia}.

The variation of the smeared Hamiltonian constraint is given by
\begin{equation}\label{Variation of Ham. const.}
\begin{split}
\delta \mathcal{H}[\tilde{N}] =& \int_\Sigma d^3x\; \tilde{N} \left(\epsilon_{ijk} \delta F^i_{ab}E^a_j E^b_k +2 \epsilon_{ijk}F^i_{ab}\delta E^a_j E^b_k -4 (\beta^2 -s) \left[\delta K^j_{[a} K^k_{b]}E^a_j E^b_k + K^j_{[a} K^k_{b]} \delta E^a_j E^b_k \right]\right) \\
& - \beta \delta \mathcal{G}_i [\mathring{\Lambda}^i]\\
=&
 \int_\Sigma \tilde{N} \left(\epsilon_{ijk} [2\partial_a \delta A^i_b + 2 \epsilon_{ilm}A^m_b \delta A^l_a] E^a_j E^b_k +2 \epsilon_{ijk}F^i_{ab}\delta E^a_j E^b_k -4 (\beta^2 -s)K^j_{[a} K^k_{b]} \delta E^a_j E^b_k \right.\\
&\hspace{1.5cm}
\left. -4 \beta^{-1}(\beta^2 -s) \delta A^j_{[a} K^k_{b]}E^a_j E^b_k 
+4 \beta^{-1}(\beta^2 -s) \delta \Gamma^j_{[a} K^k_{b]}E^a_j E^b_k \right) - \beta \delta \mathcal{G}_i [\mathring{\Lambda}^i]\\
=&
 \int_\Sigma d^3x\; \left(-2\epsilon_{njk} \partial_a(\tilde{N} E^a_j E^e_k)\delta A^n_e + 2\tilde{N} \epsilon_{ijk} \epsilon_{inm}A^m_b \delta A^n_e E^e_j E^b_k \right.\\
&\hspace{1.5cm}
+2 \epsilon_{ink} \tilde{N} F^i_{eb} E^b_k(\delta E^e_n) -4 (\beta^2 -s)\tilde{N} K^n_{[e} K^k_{b]} E^b_k (\delta E^e_n) \\
&\hspace{1.5cm}
-2 \beta^{-1}(\beta^2 -s)\tilde{N}  K^k_{b}E^e_n E^b_k (\delta A^n_{e}) 
+2 \beta^{-1}(\beta^2 -s)\tilde{N}  K^k_{a}E^a_n E^e_k (\delta A^n_{e})\\
&\hspace{1.5cm}
\left.+4 \beta^{-1}(\beta^2 -s) \tilde{N}\delta \Gamma^j_{[a} K^k_{b]}E^a_j E^b_k \right)
+
 \oint dS_a\; 2 \tilde{N} \epsilon_{ijk} \delta A^i_b E^a_j E^b_k \\
=&
 \int_\Sigma d^3x\; (\delta A^n_e) \left(-2\epsilon_{njk} \partial_a(\tilde{N} E^a_j E^e_k) + 2\tilde{N} \epsilon_{ijk} \epsilon_{inm}A^m_b E^e_j E^b_k
-4 \beta^{-1}(\beta^2 -s)\tilde{N} K^k_{a}E^{[e}_n E^{a]}_k -\beta \epsilon_{njk} \mathring{\Lambda}^j E^e_k\right)\\
&+
 \int_\Sigma d^3x\; (\delta E^e_n) E^b_k \left(
2 \epsilon_{ink} \tilde{N} F^i_{eb}  -4 (\beta^2 -s)\tilde{N} K^n_{[e} K^k_{b]} -\beta \epsilon_{njk} \mathring{\Lambda}^j A_e^k\right) \\
&+
4 \beta^{-1}(\beta^2 -s) 
 \int_\Sigma d^3x\; 
\tilde{N}\delta \Gamma^j_{[a} K^k_{b]}E^a_j E^b_k
+
 \oint dS_a\; \left(2 \tilde{N} \epsilon_{ijk} \delta A^i_b E^a_j E^b_k  - 2 \hat{\Lambda}^i \delta E^a_i\right)
\end{split}
\end{equation}
In order to obtain $\delta_{\tilde{N}} A_a^i$ and the respective surface terms, it is necessary to express the term
$ \int_\Sigma d^3x\; \tilde{N}\delta \Gamma^j_{[a} K^k_{b]}E^a_j E^b_k $ explicitly in terms of $\delta E^a_i$. We have performed these calculations as shown below. 
\begin{equation}\label{Extra cal. 1}
\begin{split}
\int_\Sigma d^3x\; & \tilde{N}\delta \Gamma^j_{[a} K^k_{b]}E^a_j E^b_k = \frac{1}{2} \int_\Sigma d^3x\; \left( \tilde{N} \delta \Gamma^j_{a} E^a_j K^k_{b} E^b_k - \tilde{N} \delta \Gamma^i_a E^a_m K^m_d E^d_i\right)\\
&=
\frac{1}{4} \int_\Sigma d^3x\; (\delta E^e_n) \left( \epsilon^{njk} \tilde{N} K^i_a E^a_i E^b_k \left[E^l_e E^c_j  E^l_{c,b}  + E^j_e \frac{(\det(E))_{,b}}{\det(E)} \right]
+ \epsilon^{ljk} \partial_a (\tilde{N} K^i_d E^d_i E^b_k E^a_l)E^n_b E^j_e \right) \\
&
+\frac{1}{4}\int_\Sigma d^3x\; \tilde{N}  \epsilon^{ijn} (\delta E^e_n) K^m_d E^d_i \left[E^a_m E^j_{a,e}-  E^a_m E^j_{e,a}+ E^c_j E^m_{c,e} + \delta^j_m \frac{(\det(E))_{,e}}{\det(E)} \right]\\
&
- \frac{1}{4}\int_\Sigma d^3x\; \tilde{N} \epsilon^{ijk} (\delta E^e_n) E^b_k K^n_d E^d_i \left[ E^c_j E^l_e  E^l_{c,b}  + E^j_e  \frac{(\det(E))_{,b}}{\det(E)} \right] 
+ \frac{1}{4}\int_\Sigma d^3x\; \tilde{N} \epsilon^{ink} (\delta E^e_n) E^b_k K^m_d E^d_i E^m_{e,b}\\
&
+ \frac{1}{4}\int_\Sigma d^3x\;  \epsilon^{ijk} (\delta E^e_n)  E^j_e \left[\partial_b (\tilde{N} E^b_k K^m_d E^d_i E^a_m) E^n_a - \partial_a (\tilde{N} E^b_k K^m_d E^d_i E^a_m) E^n_b \right]\\
&
- \frac{1}{4}\int_\Sigma d^3x\;  \epsilon^{ijk}  (\delta E^e_n)  \left[-\partial_b(\tilde{N} E^b_k K^m_d E^d_i E^c_j) E^n_c E^m_e + \partial_b (\tilde{N} E^b_k K^j_d E^d_i) E^n_e \right]\\
&+
 \frac{1}{4} \oint dS_a \epsilon^{ijk} \tilde{N} K^n_e E^e_n E^b_k E^a_i \delta E^j_{b}
+ \frac{1}{4} \oint dS_b \; \epsilon^{ijk} \tilde{N} E^b_k K^m_d E^d_i \left[E^a_m  \delta E^j_{a} + E^c_j (\delta E^m_{c})  + \delta^j_m (E^l_d \delta E^d_l)\right] \\
& -\frac{1}{4} \oint dS_a \; \epsilon^{ijk} \tilde{N} E^b_k K^m_d E^d_i E^a_m \delta E^j_b
\end{split}
\end{equation}
where the variation of the spin connection is calculated using the relation (\ref{Spin connection}). 

By substituting (\ref{Extra cal. 1}) into (\ref{Variation of Ham. const.}) and reorganizing the terms, we can express $\delta \mathcal{H}[\tilde{N}]$ in the desired form
\begin{equation}
\delta \mathcal{H}[\tilde{N}] = \Omega(\delta_{\tilde{N}}, \delta ) + \mathcal{B}_{\tilde{N}} (\delta A^i_a, \delta E^a_i)= \frac{2}{\beta} \int_\Sigma d^3x\; \left((\delta E^a_i)(\delta_{\tilde{N}} A^i_a) - (\delta A^i_a) (\delta_{\tilde{N}} E^a_i) \right)+ \mathcal{B}_{\tilde{N}} (\delta A^i_a, \delta E^a_i)
\end{equation}
From this expression, we can read both $\delta_{\tilde{N}} A^i_a, \; \delta_{\tilde{N}} E^a_i$, as well as the boundary term $\mathcal{B}_{\tilde{N}} (\delta A^i_a, \delta E^a_i)$. 
The variation of the Ashtekar-Barbero connection under the action of the Hamiltonian constraint is
\begin{equation}\label{Var. of A under Ham Const.}
\begin{split}
\delta_{\tilde{N}} A^n_e =& \;\beta E^b_k \left(\epsilon_{ink} \tilde{N} F^i_{eb}  -2 (\beta^2 -s)\tilde{N} K^n_{[e} K^k_{b]}\right)\\
&
+ \frac{1}{2}(\beta^2 -s) 
\left(
\epsilon^{njk} \tilde{N} K^m_f E^f_m E^b_k \left[E^l_e E^c_j  E^l_{c,b}  + E^j_e \frac{(\det(E))_{,b}}{\det(E)} \right]\right.\\
&\hspace{2.3cm}
+ \epsilon^{ljk} \partial_a (\tilde{N} K^i_d E^d_i E^b_k E^a_l)E^n_b E^j_e \\
&\hspace{2.3cm}
+ \tilde{N}  \epsilon^{ijn} K^m_d E^d_i \left[E^a_m E^j_{a,e}-  E^a_m E^j_{e,a}+ E^c_j E^m_{c,e} + \delta^j_m \frac{(\det(E))_{,e}}{\det(E)} \right]\\
&\hspace{2.3cm}
-  \tilde{N} \epsilon^{ijk} E^b_k K^n_d E^d_i \left[ E^c_j E^l_e  E^l_{c,b}  + E^j_e  \frac{(\det(E))_{,b}}{\det(E)} \right] \\
&\hspace{2.3cm}
+ \tilde{N} \epsilon^{ink} E^b_k K^m_d E^d_i E^m_{e,b}\\
&\hspace{2.3cm}
+ \epsilon^{ijk} E^j_e \left[\partial_b (\tilde{N} E^b_k K^m_d E^d_i E^a_m) E^n_a - \partial_a (\tilde{N} E^b_k K^m_d E^d_i E^a_m) E^n_b \right]\\
&\hspace{2.3cm}
\left.- \epsilon^{ijk} \left[-\partial_b(\tilde{N} E^b_k K^m_d E^d_i E^c_j) E^n_c E^m_e + \partial_b (\tilde{N} E^b_k K^j_d E^d_i) E^n_e \right]\right) -\beta \epsilon_{njk} \mathring{\Lambda}^j A^k_e
\end{split}
\end{equation}
The variation of the electric field is given by
\begin{equation}\label{Var. of E under Ham Const.}
\delta_{\tilde{N}} E^e_n = \beta \epsilon_{njk}\partial_a (\tilde{N} E^a_j E^e_k) - \beta \tilde{N} \epsilon_{ijk} \epsilon_{inm} A^m_b E^e_j E^b_k + 2(\beta^2 - s) \tilde{N} K^k_a E^{[e}_n E^{a]}_k -\beta \epsilon_{njk} \mathring{\Lambda}^j E^e_k
\end{equation}
The surface term associated with the Hamiltonian constraint is
\begin{equation}\label{BT of Ham. const.}
\begin{split}
\mathcal{B}_{\tilde{N}} (\delta A^i_a, \delta E^a_i) =& \oint dS_a\; 2 \tilde{N} \epsilon_{ijk} \delta A^i_b E^a_j E^b_k + \beta^{-1}(\beta^2 -s) \oint dS_a \epsilon^{ijk} \tilde{N} K^n_e E^e_n E^b_k E^a_i \delta E^j_{b}\\
&+ \beta^{-1}(\beta^2 -s) \oint dS_b \; \epsilon^{ijk} \tilde{N} E^b_k K^m_d E^d_i \left[E^a_m  \delta E^j_{a} + E^c_j (\delta E^m_{c})  + \delta^j_m (E^l_d \delta E^d_l)\right] \\
& - \beta^{-1}(\beta^2 -s) \oint dS_a \; \epsilon^{ijk} \tilde{N} E^b_k K^m_d E^d_i E^a_m \delta E^j_b - 2 \oint dS_a\;\mathring{\Lambda}^i  \delta E^a_i 
\end{split}
\end{equation}
we will deal with the the surface term issue in section \ref{Section Asymptotic charges}.

\subsection{\textsf{Preservation under hypersurface deformations}}\label{Section Preservation}
In this section we want to show that the boundary conditions proposed in the theorem are preserved under the hypersurface deformations. To do that we use the results we obtained in the appendix \ref{App. Preservation}. We begin with equation (\ref{assumption 1}). We should make sure that this equation is preserved when one moves from one hypersurface to the next. Using the variations (\ref{Var. F r D}) and (\ref{Var. F B r}), we have 
\begin{align}\label{Var. assumption 1}
&\delta (\bar{F}^B_r \bar{\gamma}_{BD} + \bar{F}^r_D)\nonumber\\ 
=&
\bar{\gamma}_{BD} \left(\beta \left[\frac{\epsilon^{AC} }{\sqrt{\bar{\gamma}}}b \; \bar{D}_A \bar{F}^B_C + \frac{\epsilon^{CB} }{\sqrt{\bar{\gamma}}}\bar{F}^A_C (\bar{D}_A b )  + \frac{\epsilon^{AB}}{\sqrt{\bar{\gamma}}} (\bar{D}_A \tilde{f})+ b \left(\bar{G}^r_A \bar{\gamma}^{AB} - \frac{(\beta^2 - s)}{\beta} \bar{k}^B_r \right)\right]+ \mathcal{L}_Y \bar{F}^B_r - \frac{\epsilon^{BC}}{\sqrt{\bar{\gamma}}}\bar{\Lambda}_C \right)\nonumber \\
&+  \beta \left(\bar{\gamma}_{AD} \frac{\epsilon^{AB}}{\sqrt{\bar{\gamma}}} \left[
b \bar{D}_B \bar{F}^r_r 
-b (\bar{\gamma}_{BC} \bar{F}^C_r - \bar{F}^r_B)
+ \bar{F}^C_B (\bar{D}_C b) 
+ (\bar{D}_B \tilde{f})\right]
+ b \; \bar{G}^A_r \bar{\gamma}_{AD} - \frac{(\beta^2 - s)}{\beta} b \; \bar{k}^r_D\right)\nonumber \\
& - \partial_D W + I_D + \mathcal{L}_Y \bar{F}^r_D + \frac{\epsilon^{AB}}{\sqrt{\bar{\gamma}}}\bar{\gamma}_{AD} \bar{\Lambda}_B \nonumber \\
=&
\beta \; b \left[\frac{\epsilon^{AC} }{\sqrt{\bar{\gamma}}} \left(\bar{\gamma}_{BD} \; \bar{D}_A \bar{F}^B_C + \bar{\gamma}_{AD} \left[\bar{D}_C \bar{F}^r_r - 2 \bar{\gamma}_{BC} \bar{F}^B_r\right] \right) + \bar{G}^r_D + \bar{G}^A_r \bar{\gamma}_{AD} - \frac{(\beta^2 - s)}{\beta}(\bar{k}^r_D + \bar{\gamma}_{BD} \bar{k}^B_r) \right] \nonumber \\
& - \partial_D W + I_D + \mathcal{L}_Y (\bar{\gamma}_{BD} \bar{F}^B_r + \bar{F}^r_D)
\end{align}
In order to make sure that (\ref{assumption 1}) is preserved we have to fix $I^A$ by the equation
\begin{align}\label{Fixed I}
I_D =& \partial_D W -\beta \; b \left[\frac{\epsilon^{AC} }{\sqrt{\bar{\gamma}}} \left(\bar{\gamma}_{BD} \; \bar{D}_A \bar{F}^B_C + \bar{\gamma}_{AD} \left[\bar{D}_C \bar{F}^r_r - 2 \bar{\gamma}_{BC} \bar{F}^B_r\right] \right) + \bar{G}^r_D + \bar{G}^A_r \bar{\gamma}_{AD} - \frac{(\beta^2 - s)}{\beta}(\bar{k}^r_D + \bar{\gamma}_{BD} \bar{k}^B_r) \right]\nonumber\\
=&
\partial_D W -\beta \; b \left[ \frac{s}{\beta}(\bar{k}^r_D + \bar{\gamma}_{BD} \bar{k}^B_r) - \frac{\epsilon^{AC} }{\sqrt{\bar{\gamma}}}  \bar{\gamma}_{AD} \bar{\gamma}_{BC} \bar{F}^B_r  \right]
\end{align}
In fact inserting (\ref{Fixed I}) in (\ref{Var. assumption 1}) leads to $\delta (\bar{F}^B_r \bar{\gamma}_{BD} + \bar{F}^r_D)=0$. 

In order to show that the equation (\ref{assumption 2}) is preserved we use the equations (\ref{Var. F B D}) to get
\begin{align}\label{Var. assumption 2}
\bar{\gamma}_{B[E} \delta \bar{F}^B_{D]} =& \beta\left[\frac{1}{\sqrt{\bar{\gamma}}} \bar{\gamma}_{B[E} \bar{\gamma}_{D]C}  \left(\epsilon^{CA} b \bar{D}_A \bar{F}^B_r - \epsilon^{CB} \bar{F}^A_r (\bar{D}_A b)\right) + b \; \bar{G}^C_{[E}  \bar{\gamma}_{D]C} \right]\nonumber\\
&- \bar{\gamma}_{B[E} \bar{D}_{D]} I^B  + \mathcal{L}_Y (\bar{\gamma}_{B[E}  \bar{F}^B_{D]} ) + \frac{\epsilon^{BC}}{\sqrt{\bar{\gamma}}}\bar{\Lambda}^r \bar{\gamma}_{B[E} \bar{\gamma}_{D]C}\nonumber\\
&=
\bar{\gamma}_{B[E} \bar{\gamma}_{D]C} \left[ \beta \left( \epsilon^{CA} b \bar{D}_A \bar{F}^B_r - \epsilon^{CB} \bar{F}^A_r (\bar{D}_A b)  + b \bar{G}^C_A \bar{\gamma}^{AB}\right)  - \bar{D}^C I^B  + \frac{\epsilon^{BC}}{\sqrt{\bar{\gamma}}} \bar{\Lambda}^r \right]  + \mathcal{L}_Y (\bar{\gamma}_{B[E}  \bar{F}^B_{D]} ) 
\end{align}
One can solve the equation $\bar{\gamma}_{B[E} \delta \bar{F}^B_{D]}=0$ for $\bar{\Lambda}^r$ in the following way
\begin{align}\label{bar lambda r fixed}
2 \bar{\Lambda}^r =& \sqrt{\bar{\gamma}} \; \epsilon_{BC} (\bar{D}^C I^B) - \beta \left[2 \bar{F}^A_r (\bar{D}_A b) - b \bar{D}_A \bar{F}^A_r + b \sqrt{\bar{\gamma}} \; \epsilon_{BC} \bar{G}^C_A \bar{\gamma}^{AB} \right]\nonumber \\
=&
\sqrt{\bar{\gamma}} \; (\frac{1}{\bar{\gamma}} \epsilon^{DA}\bar{\gamma}_{DB} \bar{\gamma}_{AC})(\bar{D}^C I^B) - \beta \left[2 \bar{F}^A_r (\bar{D}_A b) - b \bar{D}_A \bar{F}^A_r + b \sqrt{\bar{\gamma}} \; (\frac{1}{\bar{\gamma}} \epsilon^{DE}\bar{\gamma}_{DB} \bar{\gamma}_{EC}) \bar{G}^C_A \bar{\gamma}^{AB} \right]\nonumber \\
=&
 \frac{1}{\sqrt{\bar{\gamma}} } \epsilon^{DA} (\bar{D}_A I_D) - \beta \left[2 \bar{F}^A_r (\bar{D}_A b) - b \bar{D}_A \bar{F}^A_r + b  \; \frac{1}{\sqrt{\bar{\gamma}}} \epsilon^{AB} \bar{\gamma}_{BC} \bar{G}^C_A  \right]\nonumber \\
=&
\frac{\epsilon^{DA} }{\sqrt{\bar{\gamma}} } \bar{D}_A \left( \partial_D W -\beta \; b \left[ \frac{s}{\beta}(\bar{k}^r_D + \bar{\gamma}_{BD} \bar{k}^B_r) - \frac{\epsilon^{EC} }{\sqrt{\bar{\gamma}}}  \bar{\gamma}_{ED} \bar{\gamma}_{BC} \bar{F}^B_r  \right] \right)
- \beta \left[2 \bar{F}^A_r (\bar{D}_A b) - b \bar{D}_A \bar{F}^A_r + b  \; \frac{\epsilon^{AB}}{\sqrt{\bar{\gamma}}}  \bar{\gamma}_{BC} \bar{G}^C_A  \right]\nonumber \\
=&
-s \; \frac{\epsilon^{DA} }{\sqrt{\bar{\gamma}} } \bar{D}_A \left( b\;   [\bar{k}^r_D + \bar{\gamma}_{BD} \bar{k}^B_r]  \right)
- \beta \left[ \bar{F}^A_r (\bar{D}_A b) - 2 b \bar{D}_A \bar{F}^A_r + b  \; \frac{\epsilon^{AB}}{\sqrt{\bar{\gamma}}}  \bar{\gamma}_{BC} \bar{G}^C_A  \right]
\end{align}  
Plugging (\ref{bar lambda r fixed}) into (\ref{Var. assumption 2}) ensures that $\delta (\bar{\gamma}_{B[E} \bar{F}^B_{D]})=0$.

Using the equations (\ref{Var. F r r}) and (\ref{Var. F B D}), we compute the variation of $\bar{F}^r_r - \bar{F}^A_A$ that is supposed to be even.
\begin{align*}
\delta (\bar{F}^r_r - \bar{F}^A_A)
=&
\beta \left(\frac{\epsilon^{AB}}{\sqrt{\bar{\gamma}}} b \bar{D}_A \bar{F}^r_B - \frac{\epsilon^{AB}}{\sqrt{\bar{\gamma}}}\bar{\gamma}_{AC} b \bar{F}^C_B 
- b \bar{G}^A_A + \frac{(\beta^2 - s)}{\beta} b  \bar{k}^A_A \right)
+
2W + \bar{D}_A I^A + \mathcal{L}_Y \bar{F}^r_r\\
&-
\beta\left[\frac{\bar{\gamma}_{CB}}{\sqrt{\bar{\gamma}}} \left(\epsilon^{CA} b \bar{D}_A \bar{F}^B_r\right) + b\;(-\bar{G}^A_A  - 2 \bar{G}^r_r)
+ \frac{(\beta^2 - s)}{\beta}  b\;  (2\bar{k}^r_r + \bar{k}^B_B )\right]
+ 2 W  + (\bar{D}_A I^A) + \mathcal{L}_Y \bar{F}^B_B\\
=&
\beta \left( 2 b \bar{G}^r_r - \frac{(\beta^2 - s)}{\beta^2} 2 b \bar{G}^r_r - \frac{\epsilon^{AB}}{\sqrt{\bar{\gamma}}}\bar{\gamma}_{AC} b \bar{F}^C_B 
\right)
+ \mathcal{L}_Y (\bar{F}^r_r- \bar{F}^B_B)\\
=&
\beta \left( 2 b \bar{G}^r_r - \frac{(\beta^2 - s)}{\beta^2} 2 b \bar{G}^r_r 
\right)
+ \mathcal{L}_Y (\bar{F}^r_r- \bar{F}^B_B)\\
=&
\frac{2 s}{\beta} b \; \bar{G}^r_r + \mathcal{L}_Y (\bar{F}^r_r- \bar{F}^B_B)
\end{align*}
which is even because $b \sim \text{odd}$ and $\bar{G}^r_r \sim \text{odd}$. This means that the first parity condition in (\ref{PC 3}) is preserved.

Utilizing the equation (\ref{Var. F B r}) and subsequent simplification, we obtain the following expression for $\delta \bar{F}^\theta_r$
\begin{align}
\delta \bar{F}^\theta_r &= \frac{\beta}{\sqrt{\bar{\gamma}}} \left[b \bar{D}_\theta \bar{F}^\theta_\varphi - b \bar{D}_\varphi \bar{F}^\theta_\theta - \bar{F}^\theta_\varphi (\bar{D}_\theta b) - \bar{F}^\varphi_\varphi (\bar{D}_\varphi b) + \bar{D}_\varphi \tilde{f} + b \sqrt{\bar{\gamma}} \left(\bar{G}^r_\theta - \frac{\beta^2 -s}{\beta} \bar{k}^\theta_r \right)  \right] + \mathcal{L}_Y \bar{F}^\theta_r - \frac{1}{\sqrt{\bar{\gamma}}} \bar{\Lambda}_\varphi \nonumber \\
&=
\beta b  \left(\bar{G}^r_\theta - \frac{\beta^2 -s}{\beta} \bar{k}^\theta_r \right)  - \frac{1}{\sqrt{\bar{\gamma}}} \bar{\Lambda}_\varphi + \text{even} \nonumber \\
& \sim \text{even} 
\end{align}
where by employing the parity conditions (\ref{PC 1}), (\ref{PC 2}), and (\ref{parity of W and tilde f}) in the first step, we can determine the terms with transparent even parity. Subsequently, we make use of the parity $\bar{k}^\theta_r = \beta^{-1} (\bar{G}^\theta_r - \bar{\Gamma}^\theta_r) \sim \text{odd}$ derived from (\ref{PC 4}) and the equation
\begin{equation}
\bar{\Gamma}^\theta_r = \frac{1}{2 \sqrt{\bar{\gamma}}} \left(2 \bar{\gamma} \bar{F}^\varphi_r - \bar{D}_\varphi (\bar{F}^r_r - \bar{F}^A_A) \right)
\end{equation}
Furthermore, through the cancellation of the odd parts of $\beta b \bar{G}^r_\theta$ and $- \frac{1}{\sqrt{\bar{\gamma}}} \bar{\Lambda}_\varphi$ as shown in (\ref{PC 5}) and (\ref{Lambda phi}), an even quantity remains. Thus, the parity of $\bar{F}^\theta_r$ is preserved.

A similar approach is utilized for $\delta \bar{F}^\varphi_r$. By employing the equation (\ref{Var. F B r}) and simplifying, we obtain
\begin{align}
\delta \bar{F}^\varphi_r &= \frac{\beta}{\sqrt{\bar{\gamma}}} \left[b \bar{D}_\theta \bar{F}^\varphi_\varphi - b \bar{D}_\varphi \bar{F}^\varphi_\theta + \bar{F}^\theta_\theta (\bar{D}_\theta b) + \bar{F}^\varphi_\theta (\bar{D}_\varphi b) + \bar{D}_\theta \tilde{f} + b \sqrt{\bar{\gamma}} \left(\bar{G}^r_\varphi - \frac{\beta^2 -s}{\beta} \bar{k}^\varphi_r \right)  \right] + \mathcal{L}_Y \bar{F}^\varphi_r + \frac{1}{\sqrt{\bar{\gamma}}} \bar{\Lambda}_\theta \nonumber \\
&=
\left(\bar{G}^r_\varphi - \frac{\beta^2 -s}{\beta} \bar{k}^\varphi_r \right)  + \frac{1}{\sqrt{\bar{\gamma}}} \bar{\Lambda}_\theta + \text{odd} \nonumber \\
& \sim \text{odd} 
\end{align}
In the first step, we used the parity conditions (\ref{PC 1}), (\ref{PC 2}), and (\ref{parity of W and tilde f}) to determine the terms with transparent odd parity. Next, we employed the parity $\bar{k}^\varphi_r = \beta^{-1} (\bar{G}^\varphi_r - \bar{\Gamma}^\varphi_r) \sim \text{even}$ resulting from (\ref{PC 6}) and the equation
\begin{equation}
\bar{\Gamma}^\varphi_r = \frac{1}{2 \sqrt{\bar{\gamma}}} \left(-2 \bar{\gamma} \bar{F}^\theta_r - \bar{D}_\theta (\bar{F}^r_r - \bar{F}^A_A) \right)
\end{equation}
Additionally, the cancellation of the even parts of $\beta b \bar{G}^r_\theta$ and $- \frac{1}{\sqrt{\bar{\gamma}}} \bar{\Lambda}_\varphi$ is evident from (\ref{PC 5}) and (\ref{Lambda phi}), resulting in an odd quantity. Consequently, the parity of $\bar{F}^\varphi_r$ is preserved.

Furthermore, considering that the parities of $\bar{F}^\theta_r$ and $\bar{F}^\varphi_r$ are both preserved,
and due to the fact that the equation (\ref{assumption 1}) remains unchanged under hypersurface deformations, we can conclude that the parities of $\bar{F}_\theta^r$ and $\bar{F}_\varphi^r$ are also preserved. This result can also be confirmed directly from (\ref{Var. F B r}).

By using the equation (\ref{Var. F B D}), one can derive $\delta \bar{F}^\theta\theta$ and $\delta \bar{F}^\varphi\varphi$, which can be simplified as
\begin{align} 
\delta \bar{F}^\theta_\theta
&=
\beta\left[\frac{1}{\sqrt{\bar{\gamma}}}  b (\bar{D}_\varphi \bar{F}^\theta_r) - b\;(\bar{G}^r_r + \bar{G}^\varphi_\varphi)
+ \frac{(\beta^2 - s)}{\beta^2}  b\;  (\bar{G}^r_r + \bar{G}^\varphi_\varphi - \bar{\Gamma}^\varphi_\varphi )\right]
+ W  +  \bar{D}_\varphi I^\varphi + \mathcal{L}_Y \bar{F}^\theta_\theta \nonumber\\
& \sim \text{odd}
\end{align}
\begin{align} 
\delta \bar{F}^\varphi_\varphi
&=
\beta\left[- \sqrt{\bar{\gamma}}  b (\bar{D}_\theta \bar{F}^\varphi_r)  - b\;(\bar{G}^r_r + \bar{G}^\theta_\theta)
+ \frac{(\beta^2 - s)}{\beta^2}  b\;  (\bar{G}^r_r + \bar{G}^\theta_\theta - \bar{\Gamma}^\theta_\theta)\right]
+ W  + \bar{D}_\theta I^\theta + \mathcal{L}_Y \bar{F}^\varphi_\varphi \nonumber\\
& \sim \text{odd}
\end{align}
It is evident that both $\delta \bar{F}^\theta\theta$ and $\delta \bar{F}^\varphi\varphi$ are odd, as can be verified by the fact that $I^\theta \sim \text{even}$, $I^\varphi \sim \text{odd}$, and  
\begin{equation}
\bar{\Gamma}^\varphi_\varphi = \frac{1}{\sqrt{\bar{\gamma}}} \bar{D}_\varphi \bar{F}^\theta_r \sim \text{even}, \;\;\;\;\; \bar{\Gamma}^\theta_\theta = - \sqrt{\bar{\gamma}} \bar{D}_\theta \bar{F}^\varphi_r \sim \text{even}
\end{equation}
Consequently, the parities of $\bar{F}^\theta\theta$ and $\bar{F}^\varphi\varphi$ are maintained under deformations of hypersurfaces.

The off-diagonal component of $\delta \bar{F}^A_B$ can be expressed as follows in a simplified manner
\begin{align} 
\delta \bar{F}^\theta_\varphi
&=
\beta\left[\sqrt{\bar{\gamma}} \left(- b \bar{D}_\theta \bar{F}^\theta_r + \bar{F}^A_r (\bar{D}_A b)\right) + b\;\bar{\gamma} \bar{G}^\varphi_\theta
- \frac{(\beta^2 - s)}{\beta}  b\;  \bar{k}^\theta_\varphi \right] - \bar{D}_\varphi I^\theta  + \mathcal{L}_Y \bar{F}^\theta_\varphi + \sqrt{\bar{\gamma}}\bar{\Lambda}^r \nonumber\\
&=
\beta b\;\bar{\gamma} \bar{G}^\varphi_\theta + \sqrt{\bar{\gamma}}\bar{\Lambda}^r + \text{even} \nonumber\\
&\sim \text{even}
\end{align}
We arrived at this result by utilizing the fact that $\bar{k}^\theta_\varphi \sim \text{odd}$ and the observation that from (\ref{Lambdar}), it can be deduced that 
\begin{equation}
\bar{\Lambda}^r = \text{even} - \beta \frac{b}{2} (\bar{F}^r_r- \bar{F}^A_A),
\end{equation}
Therefore, the odd components of $\beta b\;\bar{\gamma} \bar{G}^\varphi_\theta$ and $\sqrt{\bar{\gamma}}\bar{\Lambda}^r$ cancel each other out. 
Additionally, given that the parity of $\bar{F}^\theta_\varphi$ remains unchanged and equation (\ref{assumption 2}) is unaffected by deformations in hypersurfaces, it can be concluded that the parity of $\bar{F}^\varphi_\theta$ is likewise preserved.

Based on the appearance of equation (\ref{Var. G r r}), it is not possible to ensure that it is odd. Therefore, we proceed to apply slight adjustments in order to transform it into a more appropriate format, facilitating the determination of its parity. The subsequent calculation outlines the procedure employed for this purpose.
\begin{align}
\delta \bar{G}^r_r =& -\beta \frac{\epsilon^{AC}}{\sqrt{\bar{\gamma}}} \bar{\gamma}_{BC} \left( \bar{D}_A (b \bar{G}^B_r) + b \bar{G}^B_A  \right) + \frac{\beta^2 -s}{2 \sqrt{\bar{\gamma}}} \bar{D}_A (b\; \epsilon^{AB} \bar{\pi}^r_B)  + \mathcal{L}_Y \bar{G}^r_r  + \bar{\Lambda}^r \nonumber \\
=&
 -\beta \frac{\epsilon^{AC}}{\sqrt{\bar{\gamma}}} \bar{\gamma}_{BC} \bar{D}_A \left( b \left[-\frac{\epsilon^{DB}}{2 \sqrt{\bar{\gamma}}} \left( \bar{\gamma}_{DE} \bar{F}^E_r - \bar{D}_D (\bar{F}^r_r - \bar{F^E_E}) \right)+ \beta \frac{1}{2 \sqrt{\bar{\gamma}}} \bar{\pi}^{r B}  \right]   \right)  -\beta b \frac{\epsilon^{AC}}{\sqrt{\bar{\gamma}}} \bar{\gamma}_{BC} \bar{G}^B_A \nonumber \\
 &+ 
 \frac{\beta^2 -s}{2 \bar{\gamma}} \bar{D}_A (b\; \epsilon^{AB} \bar{\pi}^r_B) + \mathcal{L}_Y \bar{G}^r_r  - \frac{s}{2 \bar{\gamma}} \epsilon^{D A} \bar{D}_A (b \bar{\pi}^r_D)+ \beta \left(b (\bar{D}_A \bar{F}^A_r) - \frac{1}{2} \bar{F}^A_r (\partial_A b) - \frac{1}{2} b \frac{\epsilon^{AC}}{\sqrt{\bar{\gamma}}} \bar{\gamma}_{BC}   \bar{G}^B_A  \right)\nonumber \\
 =&
 \frac{\beta}{2}   \bar{\gamma}^{AD} \bar{D}_A \left( b \left[ \bar{\gamma}_{DE} \bar{F}^E_r - \bar{D}_D (\bar{F}^r_r - \bar{F^E_E}) \right]   \right)  
+ \mathcal{L}_Y \bar{G}^r_r  + \beta \left(b (\bar{D}_A \bar{F}^A_r) - \frac{1}{2} \bar{F}^A_r (\partial_A b) - \frac{3}{2} b \frac{\epsilon^{AC}}{\sqrt{\bar{\gamma}}} \bar{\gamma}_{BC}   \bar{G}^B_A  \right) \nonumber\\
=&
\frac{3}{2}\beta b (\bar{D}_A \bar{F}^A_r) - \frac{\beta}{2} (\bar{D}_A b) [\bar{D}^A (\bar{F}^r_r - \bar{F}^B_B)] - \frac{\beta}{2} b \bar{D}_A \bar{D}^A (\bar{F}^r_r - \bar{F}^B_B) - \frac{3}{2} \beta b \frac{\epsilon^{AC}}{\sqrt{\bar{\gamma}}} \bar{\gamma}_{BC}   \bar{G}^B_A + \mathcal{L}_Y \bar{G}^r_r \nonumber\\
=&
- \frac{3}{2}\beta b (\bar{F}^r_r - \bar{F}^B_B) - \frac{\beta}{2} (\bar{D}_A b) [\bar{D}^A (\bar{F}^r_r - \bar{F}^B_B)] - \frac{\beta}{2} b \bar{D}_A \bar{D}^A (\bar{F}^r_r - \bar{F}^B_B) + \mathcal{L}_Y \bar{G}^r_r + \frac{3}{2}\beta b \frac{\bar{\mathcal{G}}^r}{\sqrt{\bar{\gamma}}}\nonumber\\
 \sim& \; \text{odd}
\end{align}
The odd parity of $\delta \bar{G}^r_r$ arises due to the fact that the first four terms in the last line of the above calculation are odd and $\bar{\mathcal{G}}^r =0$ is taken into consideration. Consequently, the parity of $\bar{G}^r_r$ remains unchanged.

The remaining parity conditions are addressed in a similar way using the equations provided in in appendix \ref{App. Preservation}. The variations of the remaining variables with definite parity conditions are presented below in a row.

\begin{align}
\delta \left( \bar{G}^r_r + \bar{G}^\theta_\theta \right)
=&
 \beta \sqrt{\bar{\gamma}} \left( \bar{D}_\varphi [b (\bar{G}^\theta_r)_{\text{odd}}] -\bar{D}_\theta [b (\bar{G}^\varphi_r)_{\text{even}} ] \right) + \frac{\beta^2 -s}{2 \sqrt{\bar{\gamma}}} \left(\bar{D}_\theta (b [\bar{k}^r_\varphi + \bar{\gamma}\bar{k}^\varphi_r]) - \bar{D}_\varphi (b [\bar{k}^r_\theta + \bar{k}^\theta_r]) \right) \nonumber  \\
&+ 
\beta \frac{b}{\sqrt{\bar{\gamma}}}\left(- \bar{\gamma} \bar{D}_\theta (\bar{G}^\varphi_r)_{\text{even}} + \bar{D}_\theta (\bar{G}^r_\varphi)_{\text{even}} - \bar{D}_\varphi (\bar{G}^r_\theta)_{\text{odd}} \right) - \beta \frac{1}{\sqrt{\bar{\gamma}}} \left((\partial_\varphi b) (\bar{G}^r_\theta)_{\text{odd}}  \right) \nonumber  \\
& + 
\frac{\beta^2 -s}{2 \sqrt{\bar{\gamma}}} \left(\bar{D}_\varphi (b \bar{k}^\theta_r) + b \bar{\gamma} \bar{k}^\varphi_\theta - b \bar{k}^\theta_\varphi - \bar{D}_\theta (b [\bar{k}^r_\varphi - \bar{\gamma}\bar{k}^\varphi_r]) + \bar{D}_\varphi (b [\bar{k}^r_\theta - \bar{k}^\theta_r]) \right) \nonumber  \\
& 
+ \mathcal{L}_Y \left( \bar{G}^r_r + \bar{G}^\theta_\theta \right) - \bar{D}_\theta (\bar{\Lambda}^\theta)_{\text{odd}} \nonumber \\
\sim & \text{even}
\end{align}

\begin{align}
\delta \left( \bar{G}^r_r + \bar{G}^\varphi_\varphi \right) 
=&
 \beta \sqrt{\bar{\gamma}} \left(\bar{D}_\varphi [ b (\bar{G}^\theta_r)_{\text{odd}} ] -\bar{D}_\theta [b (\bar{G}^\varphi_r)_{\text{even}} ]\right) + \frac{\beta^2 -s}{2 \sqrt{\bar{\gamma}}} \left(\bar{D}_\theta (b [\bar{k}^r_\varphi + \bar{\gamma}\bar{k}^\varphi_r]) - \bar{D}_\varphi (b [\bar{k}^r_\theta + \bar{k}^\theta_r]) \right) \nonumber  \\
&+ 
\beta \frac{b}{\sqrt{\bar{\gamma}}}\left(\bar{\gamma} \bar{D}_\varphi (\bar{G}^\theta_r)_{\text{odd}} -  \bar{D}_\varphi (\bar{G}^r_\theta)_{\text{odd}}  +  \bar{D}_\theta (\bar{G}^r_\varphi)_{\text{even}} \right)+ \beta \frac{1}{\sqrt{\bar{\gamma}}} \left((\partial_\theta b) (\bar{G}^r_\varphi)_{\text{even}}  \right).\nonumber  \\
& + \frac{\beta^2 -s}{2 \sqrt{\bar{\gamma}}} \left(-2 \bar{\gamma} \bar{D}_\theta (b \bar{k}^\varphi_r) 
 +  b \bar{\gamma} \bar{k}^\varphi_\theta  - b \bar{k}^\theta_\varphi   - \bar{D}_\theta (b [\bar{k}^r_\varphi - \bar{\gamma} \bar{k}^\varphi_r]) + \bar{D}_\varphi (b [\bar{k}^r_\theta - \bar{k}^\theta_r]) \right) \nonumber  \\
& 
+ \mathcal{L}_Y \left( \bar{G}^r_r + \bar{G}^\varphi_\varphi \right)  - \bar{D}_\varphi (\bar{\Lambda}^\varphi)_{\text{even}} \nonumber \\
\sim & \text{even}
\end{align}

\begin{align}
&\delta \left(\bar{G}^\theta_r + \frac{1}{2 \sqrt{\bar{\gamma}}} \bar{D}_\varphi (\bar{F}^r_r - \bar{F}^B_B) \right) \nonumber \\
& =
\beta \frac{1}{\sqrt{\bar{\gamma}}} \left( 2 b \bar{\gamma} (\bar{G}^\varphi_r)_{\text{even}} -  b (\bar{G}^r_\varphi)_{\text{even}} \right)
- (\beta^2 -s)  \sqrt{\bar{\gamma}}  b\;  \bar{k}^\varphi_r  + \mathcal{L}_Y \left(\bar{G}^\theta_r + \frac{1}{2 \sqrt{\bar{\gamma}}} \bar{D}_\varphi (\bar{F}^r_r - \bar{F}^B_B) \right)+ (\bar{\Lambda}^\theta)_{\text{odd}} \nonumber \\
& \sim \text{odd}
\end{align}

\begin{align}
&\delta \left(\bar{G}_\theta^r + \frac{1}{2 \sqrt{\bar{\gamma}}} \bar{D}_\varphi (\bar{F}^r_r - \bar{F}^B_B) \right)\nonumber \\
& =
-\beta \frac{1}{\sqrt{\bar{\gamma}}} \left(b (\bar{G}^r_\varphi)_{\text{even}} + b \bar{D}_\theta (\bar{G}^\theta_\varphi)_{\text{odd}} - \bar{D}_\varphi (b (\bar{G}^\theta_\theta)_{\text{even}}) + (\partial_\theta b) \bar{\gamma} (\bar{G}^\varphi_\theta)_{\text{odd}} \right) \nonumber  \\
&\hspace{0.5cm} 
+ \frac{\beta^2 -s}{2 \sqrt{\bar{\gamma}}} \left[2b \; \bar{k}^r_\varphi 
+2 \bar{D}_\theta (b \;\bar{k}^\theta_\varphi ) - 2 \beta^{-1} \bar{D}_\varphi \left(b [(\bar{G}^\theta_\theta)_{\text{even}} + \sqrt{\bar{\gamma}} (\bar{D}_\theta \bar{F}^\varphi_r)] \right) \right] \nonumber  \\
&\hspace{0.5cm} 
 - \frac{1}{2} \bar{D}_\theta \left(-s \; \frac{\epsilon^{DA} }{\sqrt{\bar{\gamma}} } \bar{D}_A \left( b\;   [\bar{k}^r_D + \bar{\gamma}_{BD} \bar{k}^B_r]  \right)
- \beta \left[ \bar{F}^A_r (\bar{D}_A b) - b \bar{D}_A \bar{F}^A_r \right] \right) \nonumber \\
&\hspace{0.5cm} 
+ \mathcal{L}_Y \left(\bar{G}^\theta_r + \frac{1}{2 \sqrt{\bar{\gamma}}} \bar{D}_\varphi (\bar{F}^r_r - \bar{F}^B_B) \right)+ (\bar{\Lambda}^\theta)_{\text{odd}} \nonumber\\
& \sim \text{odd}
\end{align}

\begin{align}
&\delta \left(\bar{G}^\varphi_r - \frac{1}{2 \sqrt{\bar{\gamma}}} \bar{D}_\theta (\bar{F}^r_r - \bar{F}^B_B) \right)\nonumber \\
& =
-\beta \frac{b}{\sqrt{\bar{\gamma}}}\left( 2 (\bar{G}^\theta_r)_{\text{odd}} - (\bar{G}^r_\theta)_{\text{odd}} \right)+ \frac{\beta^2 -s}{\sqrt{\bar{\gamma}}}  b \bar{k}^\theta_r 
+ \mathcal{L}_Y \left(\bar{G}^\varphi_r - \frac{1}{2 \sqrt{\bar{\gamma}}} \bar{D}_\theta (\bar{F}^r_r - \bar{F}^B_B) \right)  + (\bar{\Lambda}^\varphi)_{\text{even}} \nonumber \\
& \sim \text{even}
\end{align}

\begin{align}
&\delta \left(\bar{G}_\varphi^r - \frac{\sqrt{\bar{\gamma}}}{2 } \bar{D}_\theta (\bar{F}^r_r - \bar{F}^B_B) \right) \nonumber  \\
& =
 -\beta  \sqrt{\bar{\gamma}} \left(- b (\bar{G}^r_\theta)_{\text{odd}}  - b \bar{D}_\varphi (\bar{G}^\varphi_\theta)_{\text{odd}}  + \bar{D}_\theta (b (\bar{G}^\varphi_\varphi)_{\text{even}}) \right)
 + (\beta^2 -s) \sqrt{\bar{\gamma}} \left(\bar{D}_\theta (\beta^{-1} b [(\bar{G}^\varphi_\varphi)_{\text{even}}]) - b \bar{k}^r_\theta 
-  \bar{D}_\varphi (b \bar{k}^\varphi_\theta) \right) \nonumber  \\
& \hspace{0.5cm} 
+  \frac{\beta}{\sqrt{\bar{\gamma}}} (\partial_\varphi b) (\bar{G}_\varphi^\theta)_{\text{odd}}
 - \frac{1}{2} \bar{D}_\varphi \left(-s \; \frac{\epsilon^{DA} }{\sqrt{\bar{\gamma}} } \bar{D}_A \left( b\;   [\bar{k}^r_D + \bar{\gamma}_{BD} \bar{k}^B_r]  \right)
- \beta \left[ \bar{F}^A_r (\bar{D}_A b) - b \bar{D}_A \bar{F}^A_r \right] \right) \nonumber  \\
& \hspace{0.5cm} 
+ \mathcal{L}_Y \left(\bar{G}_\varphi^r - \frac{\sqrt{\bar{\gamma}}}{2 } \bar{D}_\theta (\bar{F}^r_r - \bar{F}^B_B) \right)+  (\bar{\Lambda}_\varphi)_{\text{even}}\nonumber \\
& \sim \text{even}
\end{align}

\begin{align}
&\delta \left(\bar{G}^\theta_\varphi + \frac{\sqrt{\bar{\gamma}}}{2} (\bar{F}^r_r-\bar{F}^B_B) \right) \nonumber  \\
&=
-\beta b \sqrt{\bar{\gamma}}\bar{D}_\varphi (\bar{G}^\varphi_r)_{\text{even}}
+ (\beta^2 -s)\sqrt{\bar{\gamma}} \bar{D}_\varphi ( b \bar{k}^\varphi_r) 
- \frac{\beta }{\sqrt{\bar{\gamma}}} (\partial_\varphi b)(\bar{G}^r_\varphi)_{\text{even}} 
- \bar{D}_\varphi (\bar{\Lambda}^\theta)_{\text{odd}} 
+ \mathcal{L}_Y \left(\bar{G}^\theta_\varphi + \frac{\sqrt{\bar{\gamma}}}{2} (\bar{F}^r_r-\bar{F}^B_B) \right)\nonumber \\
& \sim \text{odd}
\end{align}

\begin{align}
&\delta \left(\bar{G}_\theta^\varphi - \frac{1}{2 \sqrt{\bar{\gamma}}} (\bar{F}^r_r-\bar{F}^B_B) \right)\nonumber \\
&=
\beta \frac{b}{\sqrt{\bar{\gamma}}} \bar{D}_\theta (\bar{G}^\theta_r)_{\text{odd}} - \frac{\beta^2 -s}{\sqrt{\bar{\gamma}}}  \bar{D}_\theta ( b \bar{k}^\theta_r)  + \frac{\beta }{\sqrt{\bar{\gamma}}} (\partial_\theta b) (\bar{G}^r_\theta)_{\text{odd}} 
- \bar{D}_\theta (\bar{\Lambda}^\varphi)_{\text{even}} 
+ \mathcal{L}_Y \left(\bar{G}_\theta^\varphi - \frac{1}{2 \sqrt{\bar{\gamma}}} (\bar{F}^r_r-\bar{F}^B_B)  \right)\nonumber \\
& \sim \text{odd}
\end{align}
where to obtain the parities we have used the parity conditions in the theorem.

This demonstrates that the newly established boundary conditions remain intact when subjected to deformations of hypersurfaces. Hence, the requirement (ii) is satisfied. The sole matter that still requires examination pertains to the derived surface terms as described in section \ref{Section Constraints}, consequently giving rise to the asymptotic charges. This is the task that we will undertake in the subsequent section.

\subsection{\textsf{Asymptotic charges}}\label{Section Asymptotic charges}
Thus far, we have shown that the new boundary conditions ensure their invariance under deformations of hypersurfaces as well as the well-definedness of the symplectic structure. We will now investigate the circumstances under which the canonical generators of the asymptotic symmetries remain well-defined when the parity conditions introduced in the theorem are applied. In other words, we must first examine whether the boundary terms are finite, and then determine if they are exact.

Our objective is to prove that the bulk portion of the generators, defined by the smeared constraints $\int d^3x (\mathcal{H}[\tilde{N}]+ \mathcal{H}_a[N^a]+ \mathcal{G}[\Lambda])$, can be supplemented with appropriate surface terms that render the sum functionally differentiable.
\begin{equation}
\begin{split}
&\delta_{(\mathbf{N}, \Lambda)} \left(\int d^3x (\mathcal{H}[\tilde{N}]+ \mathcal{H}_a[N^a]+ \mathcal{G}[\Lambda])\right)  \\
&\hspace{3cm} =\frac{2}{\beta} \int_\Sigma d^3x\; \left((\delta E^a_i)(\delta_{(\mathbf{N}, \Lambda)} A^i_a) - (\delta A^i_a) (\delta_{(\mathbf{N}, \Lambda)} E^a_i) \right) + \mathcal{B}_{(\mathbf{N}, \Lambda)} (\delta A^i_a, \delta E^a_i)
\end{split}
\end{equation}
where 
\begin{equation}
\mathcal{B}_{(\mathbf{N}, \Lambda)} (\delta A^i_a, \delta E^a_i) := \mathcal{B}_{\tilde{N}} (\delta A^i_a, \delta E^a_i) + \mathcal{B}_{\vec{N}} (\delta A^i_a, \delta E^a_i) + \mathcal{B}_{\Lambda} (\delta A^i_a, \delta E^a_i) 
\end{equation}
and the explicit expressions of the boundary terms are given in (\ref{Variation of Ham. const.}), (\ref{BT of diff. const.}) and (\ref{BT of Guass. const.}) respectively. 

By employing the asymptotic expansion and gathering all divergent and finite terms, we derive the following expression
\begin{equation}\label{Asymptotic Charge Ashtekar}
\begin{split}
\mathcal{B}_{(\mathbf{N}, \Lambda)}& (\delta A^i_a, \delta E^a_i)\\
 =& \; 2 r \oint d\sigma \left[(b\epsilon^{AB} \bar{\gamma}_{BC} \delta \bar{G}^C_A - b \sqrt{\bar{\gamma}} \delta \bar{F}^r_r - \sqrt{\bar{\gamma}} (\partial_A b) \bar{\gamma}^{AB} \delta \bar{F}^r_B) 
+
\frac{s}{\beta}  \sqrt{\bar{\gamma}} Y^A (\delta \bar{G}^r_A -2 \epsilon^{BC} \bar{\gamma}_{AC} \delta \bar{F}^r_B)\right]\\
&+
\oint d\sigma \left[-2  \epsilon^{AB} \left\{b \bar{F}^r_B \delta \bar{G}^r_A - b \bar{F}^r_r \bar{\gamma}_{BC} \delta \bar{G}^C_A - b \bar{F}^C_A \bar{\gamma}_{BD} \delta \bar{G}^D_C - b  \bar{\gamma}_{BC} \delta \bar{G}^{(2)C}_A - \tilde{f} \bar{\gamma}_{BC} \delta \bar{G}^C_A  \right. \right.\\
&\hspace{1.5cm}-
\frac{\beta^2 - s}{\beta} b  \left.  \left(
 \bar{k}^D_B \bar{\gamma}_{CD} \delta \bar{F}^C_A  
- \bar{k}^r_A \delta \bar{F}^r_B 
\right) 
\right\} -2 \sqrt{\bar{\gamma}} (b \delta \bar{F}^{(2)r}_r + (\partial_A b) \bar{\gamma}^{AB} \delta \bar{F}^{(2)r}_B)\\
&\hspace{1.5cm}+
\frac{2s}{\beta} \sqrt{\bar{\gamma}} \left(- W \delta \bar{G}^A_A  + I^A \delta \bar{G}^r_A + Y^A \delta \bar{G}^{(2) r}_A  + Y^A \delta(\bar{F}^r_a  \bar{G}^a_A) -2 Y^A \frac{\epsilon^{BC}}{\sqrt{\bar{\gamma}}} \bar{\gamma}_{AC} \bar{F}^{(2)r}_B \right)\\
&\hspace{1.5cm}+\left.
\frac{2}{\beta}  \sqrt{\bar{\gamma}} \bar{\Lambda}^a \delta \bar{F}^r_a \right]
\end{split}
\end{equation}

As previously discussed in section \ref{Section H-T boundary conditions}, when working with the ADM variables, vanishing of the leading terms of the constraints were used to eliminate the divergent portion of the surface terms. However, in this case, the divergent part of the surface terms (i.e. the first line in equation (\ref{Asymptotic Charge Ashtekar})) cannot be eliminated either by employing equations (\ref{LOT radial Gauss}), (\ref{LOT angular Gauss}), (\ref{LOT radial Diff}), (\ref{LOT angular Diff}), and (\ref{LOT Ham.}), nor by applying parity conditions. This discrepancy is unexpected since the Ashtekar-Barbero variables formulation of General Relativity is expected to be equivalent to the formulation utilizing the ADM variables. In the subsequent section, we will investigate and address this fundamental difference. 

The notable aspect regarding the aforementioned surface term (\ref{Asymptotic Charge Ashtekar}) is that if we disregard asymptotic rotations and boosts (i.e. by setting $b=Y^A=0$), the divergent term is eradicated, while the corresponding charge for supertranslations remains integrable and non-zero. In other words,
\begin{equation}\label{Asymptotic Charge Ashtekar without boosts and rotations}
\begin{split}
\mathcal{B}_{(\mathbf{N}, \Lambda)}(\delta A^i_a, \delta E^a_i)\left|_{b=Y^A=0}\right.
 &= 
\oint d\sigma \left[2  \epsilon^{AB} \tilde{f} \bar{\gamma}_{BC} \delta \bar{G}^C_A 
+\frac{2s}{\beta} \sqrt{\bar{\gamma}} \left(- W \delta \bar{G}^A_A  + (\bar{D}^A W) \delta \bar{G}^r_A \right) \right]\\
&=
\oint d\sigma \left[2  \epsilon^{AB} \tilde{f} \bar{\gamma}_{BC} \delta \bar{G}^C_A 
-\frac{2s}{\beta} \sqrt{\bar{\gamma}} W \left( \delta \bar{G}^A_A  + \bar{D}^A \delta \bar{G}^r_A \right) \right]\\
&=
\delta \left(2\oint d\sigma \sqrt{\bar{\gamma}} \left[\tilde{f} (\bar{F}^r_r- \bar{F}^A_A) 
+\frac{2s}{\beta}  W  \bar{G}^r_r \right]\right)
\end{split}
\end{equation}
Hence, based on equation (\ref{definition of charge}), the charges associated with supertranslations can be expressed as
\begin{equation}\label{ST charge Ash}
\mathcal{Q}_{Supertranslation}=
- 2\oint d\sigma \sqrt{\bar{\gamma}} \left[\tilde{f} (\bar{F}^r_r- \bar{F}^A_A) 
+\frac{2s}{\beta}  W  \bar{G}^r_r \right]
\end{equation}
These charges do not vanish in general, as they represent integrals of non-trivial even functions. This outcome holds significant importance since, even in the absence of boosts and rotations in the expansion of lapse and shift, respectively, the charges linked to supertranslations become null when standard boundary conditions are employed (See equation (\ref{Simple version of Lapse and Shift}) and its accompanying explanation). Conversely, the newly proposed parity conditions in section \ref{Section Explicit form} yield non-zero supertranslation charges. Note that when $\tilde{f}$ and $W$ represent arbitrary angle-dependent even and odd functions, respectively, the non-zero supertranslation charges emerge. Additionally, the zero modes of $\tilde{f}$ and $W$ give rise to the customary charges associated with ordinary translations.

\section{\textsf{Comparison with Henneaux and Troessaert's paper}}\label{Section comparison with H-T}
In this section, we begin by transcribing the boundary conditions imposed on the Ashtekar-Barbero variables into the ADM variables. This transcription is performed in order to facilitate a comparison between the results obtained in this paper and those obtained in \cite{Henneaux}. It should be recalled, as discussed in Section \ref{Section H-T boundary conditions}, that the leading terms of canonical variables are $\bar{\lambda}, \bar{h}_{rA}, \bar{k}_{AB}$ (refer to (\ref{def. K and lambda}) for the definition of $\bar{\lambda}$ and $\bar{k}_{AB}$), along with leading terms of their conjugate momenta $\bar{p}, \bar{\pi}^{rA}, \bar{\pi}^{AB}$ (refer to (\ref{def. bar p}) for the definition of $\bar{p}$). It is assumed that $\bar{h}_{rA}$ is equal to zero, as stated in Section \ref{Section H-T boundary conditions}. 

By utilizing the relation $q_{ab}= e^i_a e^i_b = q^{-1} E^i_a E^i_b$ and performing the asymptotic expansion of both sides, we can express the variables $\bar{\lambda}, \bar{h}_{rA}, \bar{k}_{AB}$ in terms of $\bar{F}^a_b$ as 
\begin{align}
&\bar{\lambda}= \frac{1}{2} \bar{h}_{rr}= \frac{1}{2} \left(\bar{F}^a_a -2\gamma^i_r \bar{f}^i_r\right) = \frac{1}{2} \left(\bar{F}^A_A - \bar{F}^r_r \right) \sim \text{even},\label{bar lambda parity}\\
&\bar{h}_{rA}= \bar{f}^r_i \gamma^i_A + \bar{\gamma}_{AB}\bar{f}^B_i \gamma^i_r = \bar{F}^r_A + \bar{\gamma}_{AB}\bar{F}^B_r = 0, \label{h r A = 0}\\
&\bar{h}_{AB}= \bar{F}^a_a \bar{\gamma}_{AB} - \gamma^i_A \bar{f}^i_B - \gamma^i_B \bar{f}^i_A \label{h AB in terms of f}
\end{align}
Here, we have used equations (\ref{BC-Spherical}), (\ref{Asym. Exp. Inverse E}), (\ref{inverse of little f}), and (\ref{expansion of det q}). The parity of (\ref{bar lambda parity}) is inferred from (\ref{PC 3}) and equation (\ref{h r A = 0}) from (\ref{assumption 1}).

Using (\ref{h AB in terms of f}), the components of $\bar{k}_{AB}$ can be determined as
\begin{align}
\bar{k}_{AB} = \frac{1}{2} (\bar{h}_{AB}+ \bar{h}_{rr} \bar{\gamma}_{AB}) \; \Rightarrow \; &\bar{k}_{\theta \theta} = \bar{f}^\varphi_i \gamma^i_\varphi = \bar{F}^\varphi_\varphi \sim \text{odd}, \\
&\bar{k}_{\varphi \varphi} = \bar{\gamma}\bar{f}^\theta_i \gamma^i_\theta = \bar{\gamma} \bar{F}^\theta_\theta \sim \text{odd},\\
&\bar{k}_{\theta \varphi} = -\frac{1}{2}(\bar{f}^\theta_i \gamma^i_\varphi + \bar{\gamma}\bar{f}^\varphi_i \gamma^i_\theta)=  -\frac{1}{2}(\bar{F}^\theta_\varphi + \bar{\gamma}\bar{F}^\varphi_\theta)\sim \text{even}
\end{align}
The parities of these variables have been determined based on (\ref{PC 1}) and (\ref{PC 2}).

Using the relation $\pi^{ab} = 2 |\det(E)|^{-1} E^a_k E^d_k K^j_{[d} \delta^b_{c]} E^c_j$, we can express the ADM momenta in terms of the Ashtekar-Barbero variables.
\begin{align}
\pi^{rr} &= 2 |\det(E)|^{-1} E^r_k E^d_k K^j_{[d} \delta^r_{c]} E^c_j \nonumber\\
&=
|\det(E)|^{-1} E^r_k (E^d_k K^j_d E^r_j - E^r_k K^j_c E^c_j) \nonumber\\
&=
|\det(E)|^{-1} E^r_k (E^A_k K^j_A E^r_j - E^r_k K^j_A E^A_j) \nonumber\\
&=
-\sqrt{\bar{\gamma}}\; \bar{k}^j_A \gamma^A_j + O(r^{-1})
\end{align}

\begin{align}
\pi^{rA} &= 2 |\det(E)|^{-1} E^r_k E^d_k K^j_{[d} \delta^A_{c]} E^c_j\nonumber\\
&=
|\det(E)|^{-1} E^r_k (E^d_k K^j_d E^A_j - E^A_k K^j_c E^c_j)\nonumber\\
&=
|\det(E)|^{-1} E^r_k (E^A_k K^j_A E^r_j - E^r_k K^j_A E^A_j)\nonumber\\
&=
|\det(E)|^{-1} E^r_k (E^r_k K^j_r E^A_j + E^B_k K^j_B E^A_j - E^A_k K^j_r E^r_j - E^A_k K^j_B E^B_j)\nonumber\\
&=
\frac{1}{r}\sqrt{\bar{\gamma}}\; (\bar{k}^j_r \gamma^A_j + \bar{\gamma}^{AB}\bar{k}^j_B \gamma^r_j)+ O(r^{-2})
\end{align}

\begin{align}
\pi^{AB} &= 2 |\det(E)|^{-1} E^A_k E^d_k K^j_{[d} \delta^B_{c]} E^c_j \nonumber\\
&=
|\det(E)|^{-1} E^A_k (E^d_k K^j_d E^B_j - E^B_k K^j_c E^c_j)\nonumber\\
&=
|\det(E)|^{-1} E^A_k (E^r_k K^j_r E^B_j + E^C_k K^j_C E^B_j - E^B_k K^j_r E^r_j - E^B_k K^j_C E^C_j)\nonumber\\
&=
\frac{1}{r^2}\sqrt{\bar{\gamma}}\; (\bar{\gamma}^{AC}\bar{k}^j_C \gamma^B_j - \bar{\gamma}^{AB} \bar{k}^j_r \gamma^r_j - \bar{\gamma}^{AB}\bar{k}^j_C \gamma^C_j)+ O(r^{-3})
\end{align}
 From the above equations, we can easily read the leading order terms 
 \begin{align}
 &\bar{\pi}^{rr} =  -\sqrt{\bar{\gamma}}\; \bar{k}^A_A \label{bar pi rr}\\
 &\bar{\pi}^{rA} = \sqrt{\bar{\gamma}}\; (\bar{k}^A_r + \bar{\gamma}^{AB} \bar{k}^r_B) \label{bar pi rA}\\
 &\bar{\pi}^{AB} = \sqrt{\bar{\gamma}}\; (\bar{\gamma}^{AC}\bar{k}^B_C - \bar{\gamma}^{AB} \bar{k}^r_r - \bar{\gamma}^{AB}\bar{k}^C_C) \label{bar pi AB}
 \end{align}
 Now we need to express $\bar{p}, \bar{\pi}^{rA}, \bar{\pi}^{AB}$ in terms of $\bar{F}^a_b$ and $\bar{G}^a_b$. Let's start with $\bar{p}$, as defined in (\ref{def. bar p})
 \begin{align}\label{bar p in terms of G r r}
\bar{p} &= 2(\bar{\pi}^{rr}- \bar{\pi}^A_A)= 2\left(-\sqrt{\bar{\gamma}}\; \bar{k}^A_A - \sqrt{\bar{\gamma}}\; (\bar{k}^C_C - 2 \bar{k}^r_r - 2\bar{k}^C_C)\right) \nonumber \\
&= 4 \sqrt{\bar{\gamma}}\; \bar{k}^r_r 
= 4 \sqrt{\bar{\gamma}} \beta^{-1}\; (\bar{G}^r_r - \bar{\Gamma}^r_r) \nonumber \\
&= 
4 \sqrt{\bar{\gamma}} \beta^{-1}
\left[\bar{G}^r_r +\frac{1}{2} \frac{1}{\sqrt{\gamma}} \epsilon^{BA} \bar{\gamma}_{C[B}\bar{F}^C_{A]} \right]\nonumber \\
&= 
4 \sqrt{\bar{\gamma}} \beta^{-1} \bar{G}^r_r \sim \text{odd} 
 \end{align}
where in the last step we have used (\ref{assumption 2}). We have found that $\bar{p}$ has an odd parity based on the parity condition (\ref{PC 2}).

Next, we examine $\bar{\pi}^{r\theta}$ using equation (\ref{bar pi rA}) by substituting in it the explicit expressions of $\bar{\Gamma}^a_b$ from (\ref{LOT radial spin connection}) and (\ref{LOT angular spin connection}). Then, the expression for $\bar{\pi}^{r\theta}$ is 
\begin{align}
\bar{\pi}^{r\theta} &= \sqrt{\bar{\gamma}} (\bar{k}^\theta_r + \bar{\gamma}^{\theta B}\bar{k}^r_B) = \sqrt{\bar{\gamma}} (\bar{k}^\theta_r + \bar{k}^r_\theta)\nonumber\\ 
&=
\sqrt{\bar{\gamma}} \beta^{-1} \left(\bar{G}^\theta_r + \bar{G}_\theta^r - (\bar{\Gamma}^\theta_r + \bar{\Gamma}_\theta^r) \right)\nonumber\\ 
&=
\sqrt{\bar{\gamma}} \beta^{-1} \left(\bar{G}^\theta_r + \bar{G}_\theta^r - 
\left(\frac{1}{2 \sqrt{\bar{\gamma}}}\left[2 \epsilon^{\theta \varphi} \bar{\gamma}_{\varphi \varphi} \bar{F}^\varphi_r - \epsilon^{\theta \varphi} \bar{D}_\varphi (\bar{F}^r_r - \bar{F}^B_B)\right] \right. \right.\nonumber\\
&\hspace*{3.8cm}
\left.\left.
+ \frac{1}{2 \sqrt{\bar{\gamma}}}\left[2 \epsilon^{\theta \varphi} \bar{\gamma}_{\theta \theta}\bar{\gamma}_{\varphi \varphi} \bar{F}^\varphi_r - \epsilon^{\theta \varphi} \bar{\gamma}_{\theta \theta} \bar{D}_\varphi (\bar{F}^r_r + \bar{F}^B_B) + 2\epsilon^{\theta \varphi} \bar{\gamma}_{\theta \theta} \bar{D}_\varphi \bar{F}^\theta_\theta + 2\epsilon^{\varphi \theta} \bar{\gamma}_{\varphi \varphi} \bar{D}_\theta \bar{F}^\varphi_\theta \right]
\right) \right)\nonumber\\ 
&=
\sqrt{\bar{\gamma}} \beta^{-1} \left(\bar{G}^\theta_r + \bar{G}_\theta^r + \frac{1}{\sqrt{\bar{\gamma}}} \bar{D}_\varphi (\bar{F}^r_r - \bar{F}^B_B)\right) + \text{odd}\nonumber\\
& \sim \text{odd}
\end{align}
By using the parities proposed in the theorem, particularly (\ref{PC 4}) and (\ref{PC 5}), we have deduced that $\bar{\pi}^{r\theta}$ has an odd parity.

Similarly, we can express $\bar{\pi}^{r\varphi}$ in terms of $\bar{F}^a_b$ and $\bar{G}^a_b$ using the equation (\ref{bar pi rA}), (\ref{LOT radial spin connection}), (\ref{LOT angular spin connection}) and (\ref{LOT extrinsic curvature})
\begin{align}
\bar{\pi}^{r\varphi} &= \sqrt{\bar{\gamma}} (\bar{k}^\varphi_r + \bar{\gamma}^{\varphi B}\bar{k}^r_B) = \sqrt{\bar{\gamma}} (\bar{k}^\varphi_r + \bar{\gamma}^{-1}\bar{k}^r_\varphi)\nonumber\\ 
&=
\sqrt{\bar{\gamma}} \beta^{-1} \left(\bar{G}^\varphi_r + \bar{\gamma}^{-1} \bar{G}_\varphi^r - (\bar{\Gamma}^\varphi_r + \bar{\gamma}^{-1}\bar{\Gamma}_\varphi^r) \right)\nonumber\\ 
&=
\sqrt{\bar{\gamma}} \beta^{-1} \left(\bar{G}^\varphi_r + \bar{\gamma}^{-1}\bar{G}_\varphi^r - 
\left(\frac{1}{2 \sqrt{\bar{\gamma}}}\left[2 \epsilon^{\varphi \theta}  \bar{\gamma}_{\theta \theta} \bar{F}^\theta_r - \epsilon^{ \varphi \theta} \bar{D}_\theta (\bar{F}^r_r - \bar{F}^B_B)\right] \right. \right.\nonumber\\
&\hspace*{3.8cm}
\left.\left.
+ \frac{1}{2 \bar{\gamma} \sqrt{\bar{\gamma}}}\left[2 \epsilon^{\varphi \theta} \bar{\gamma}_{\varphi \varphi} \bar{\gamma}_{\theta \theta} \bar{F}^\theta_r - \epsilon^{\varphi \theta} \bar{\gamma}_{\varphi \varphi} \bar{D}_\theta (\bar{F}^r_r + \bar{F}^B_B) + 2\epsilon^{\theta \varphi} \bar{\gamma}_{\theta \theta} \bar{D}_\varphi \bar{F}^\theta_\varphi + 2\epsilon^{\varphi \theta} \bar{\gamma}_{\varphi \varphi} \bar{D}_\theta \bar{F}^\varphi_\varphi \right]
\right) \right)\nonumber\\ 
&=
\sqrt{\bar{\gamma}} \beta^{-1} \left(\bar{G}^\varphi_r + \bar{\gamma}^{-1}\bar{G}_\varphi^r - \frac{1}{\sqrt{\bar{\gamma}}} \bar{D}_\theta (\bar{F}^r_r - \bar{F}^B_B)\right) + \text{even}\nonumber\\
& \sim \text{even}
\end{align}
By utilizing the parities proposed in the theorem, specifically (\ref{PC 6}) and (\ref{PC 7}), we conclude that $\bar{\pi}^{r\varphi}$ has an even parity.

To express $\bar{\pi}^{AB}$ in terms of $\bar{F}^a_b$ and $\bar{G}^a_b$, we use the equation (\ref{bar pi AB}), (\ref{LOT extrinsic curvature}), (\ref{LOT radial spin connection}) and (\ref{LOT angular spin connection}).
\begin{align}
\bar{\pi}^\theta_\theta &= \sqrt{\bar{\gamma}} (-\bar{k}^r_r - \bar{k}^\varphi_\varphi)= - \sqrt{\bar{\gamma}} \beta^{-1}(\bar{G}^r_r + \bar{G}^\varphi_\varphi - \bar{\Gamma}^\varphi_\varphi)\nonumber\\
&=
- \sqrt{\bar{\gamma}} \beta^{-1}(\bar{G}^r_r + \bar{G}^\varphi_\varphi + \frac{1}{\sqrt{\bar{\gamma}}} \epsilon^{\varphi \theta}\bar{\gamma}_{\theta \theta} \bar{D}_\varphi \bar{F}^\theta_r)\nonumber\\
&=
- \sqrt{\bar{\gamma}} \beta^{-1}(\bar{G}^r_r + \bar{G}^\varphi_\varphi - \frac{1}{\sqrt{\bar{\gamma}}}  \bar{D}_\varphi \bar{F}^\theta_r) \sim \text{even}
\end{align}
\begin{align}
\bar{\pi}^\varphi_\varphi &= \sqrt{\bar{\gamma}} (-\bar{k}^r_r - \bar{k}^\theta_\theta)= - \sqrt{\bar{\gamma}} \beta^{-1}(\bar{G}^r_r + \bar{G}^\theta_\theta - \bar{\Gamma}^\theta_\theta)\nonumber\\
&=
- \sqrt{\bar{\gamma}} \beta^{-1}(\bar{G}^r_r + \bar{G}^\theta_\theta + \frac{1}{\sqrt{\bar{\gamma}}} \epsilon^{ \theta \varphi}\bar{\gamma}_{\varphi \varphi} \bar{D}_\theta \bar{F}^\varphi_r)\nonumber\\
&=
- \sqrt{\bar{\gamma}} \beta^{-1}(\bar{G}^r_r + \bar{G}^\theta_\theta - \frac{1}{\sqrt{\bar{\gamma}}} \bar{D}_\theta \bar{F}^\varphi_r) \sim \text{even}
\end{align}
where the assumption $\bar{\Gamma}^r_r =0$ (obtained from the relation (\ref{assumption 2})) is used in this derivation. 
The parity conditions (\ref{PC 1})-(\ref{PC 3}) are also applied to deduce that $\bar{\pi}^\theta_\theta \sim \bar{\pi}^\varphi_\varphi \sim \text{even}$.

Furthermore, the expression for $\bar{\pi}^\theta_\varphi$ and $\bar{\pi}_\theta^\varphi$ are found as
\begin{align}
\bar{\pi}^\theta_\varphi &= \sqrt{\bar{\gamma}} \bar{k}^\theta_\varphi = \sqrt{\bar{\gamma}} \beta^{-1} (\bar{G}^\theta_\varphi - \bar{\Gamma}^\theta_\varphi)\nonumber\\
&=
\sqrt{\bar{\gamma}} \beta^{-1}\left(\bar{G}^\theta_\varphi - \frac{1}{2 \sqrt{\bar{\gamma}}} \left[\epsilon^{ \theta \varphi}\bar{\gamma}_{\varphi \varphi}(\bar{F}^C_C - \bar{F}^r_r) - 2  \epsilon^{\theta \varphi}\bar{\gamma}_{\varphi \varphi}\bar{D}_\varphi \bar{F}^\varphi_r)\right] \right)\nonumber\\
&=
\sqrt{\bar{\gamma}} \beta^{-1}\left(\bar{G}^\theta_\varphi - \frac{\sqrt{\bar{\gamma}}}{2} \left[\bar{F}^C_C - \bar{F}^r_r - 2  \bar{D}_\varphi \bar{F}^\varphi_r \right]\right) \sim \text{odd}
\end{align}

\begin{align}
\bar{\pi}_\theta^\varphi &= \sqrt{\bar{\gamma}} \bar{k}_\theta^\varphi = \sqrt{\bar{\gamma}} \beta^{-1} (\bar{G}_\theta^\varphi - \bar{\Gamma}_\theta^\varphi)\nonumber\\
&=
\sqrt{\bar{\gamma}} \beta^{-1}\left(\bar{G}_\theta^\varphi - \frac{1}{2 \sqrt{\bar{\gamma}}} \left[\epsilon^{\varphi  \theta}\bar{\gamma}_{\theta \theta}(\bar{F}^C_C - \bar{F}^r_r) - 2 \epsilon^{\varphi \theta} \bar{\gamma}_{\theta \theta}\bar{D}_\theta \bar{F}^\theta_r\right] \right)\nonumber\\
&=
\sqrt{\bar{\gamma}} \beta^{-1}\left(\bar{G}_\theta^\varphi + \frac{1}{2 \sqrt{\bar{\gamma}}} \left[\bar{F}^C_C - \bar{F}^r_r - 2  \bar{D}_\theta \bar{F}^\theta_r \right]\right) \sim \text{odd}
\end{align}
The parity conditions (\ref{PC 1}), (\ref{PC 2}), (\ref{PC 8}), and (\ref{PC 9}) are utilized to conclude that $\bar{\pi}^\theta_\varphi \sim \bar{\pi}_\theta^\varphi \sim \text{odd}$.

The cumulative summation of the aforementioned calculations concludes that if the parity conditions stated in the theorem are translated to the ADM variables, the resultant parity conditions (\ref{H-T-Parity 2}) as introduced in \cite{Henneaux} are exactly obtained. The question that arises here is why, despite the application of the same parity condition to the theory, the divergence of surface terms is eliminated using the leading order terms of the constraints when working with ADM variables, but not when working with Ashtekar-Barbero variables. By examining the process of obtaining Ashtekar-Barbero variables from ADM variables, it becomes evident that in order to establish $A^i_a$ and $E^a_i$ as conjugate variables, particularly to demonstrate that the Poisson bracket between two $A^i_a$ is zero, it is necessary to prove 
\begin{equation}
\frac{\delta \Gamma^j_a (x)}{\delta E^b_k (y)}- \frac{\delta \Gamma^k_b (y)}{\delta E^a_j (x)}=0
\end{equation}
which represents the integrability condition for $\Gamma^j_a$ to possess a generating function $F$. When working with a manifold without a boundary, such an $F$ can be easily found. In addition, when working within the framework of asymptotically flat spacetimes, it is readily demonstrated that the standard boundary conditions result in a well-defined $F$. Nonetheless, when one wishes to relax the boundary conditions, matters become more complex and the well-definedness of $F$ should be regarded as an additional requirement, in addition to the three requirements necessary to propose appropriate boundary conditions. If a well-defined $F$ cannot be defined under specific boundary conditions, it indicates that the Ashtekar-Barbero variables are not canonically transformed from the ADM variables under those boundary conditions. This is manifested just in surface terms, and consequently, all analyses pertaining to the bulk, such as the well-definedness of the symplectic structure and the preservation of boundary conditions under hypersurface deformations, remain the same in both sets of variables.

The intriguing aspect of this situation is that, despite the significant disparity, the supertranslation charges derived in this paper using Ashtekar-Barbero variables, denoted as equation (\ref{ST charge Ash}), are equivalent to those presented in \cite{Henneaux}, denoted as equation (\ref{STcharge}). To demonstrate this, we will begin with equation (\ref{ST charge Ash}) and attempt to express it in terms of ADM variables. 
To elaborate, we have
\begin{align}\label{ST Ash = ST ADM}
\mathcal{Q}_{Supertranslation}&=
- 2\oint d\sigma \sqrt{\bar{\gamma}} \left[\tilde{f} (\bar{F}^r_r- \bar{F}^A_A) 
+\frac{2s}{\beta}  W  \bar{G}^r_r \right] \nonumber \\
&=
- 2\oint d\sigma \sqrt{\bar{\gamma}} \left[\tilde{f} \left(-2\bar{\lambda}\right) 
+\frac{2s}{\beta}  W  \left(\frac{\beta}{4\sqrt{\bar{\gamma}}}\;\bar{p}\right) \right]\nonumber\\
&=
\oint d\sigma \left[4 \tilde{f}  \sqrt{\bar{\gamma}} \bar{\lambda} - s W \bar{p} \right]
\end{align}
In arriving at equation (\ref{ST Ash = ST ADM}), we have utilized equations (\ref{bar lambda parity}) and (\ref{bar p in terms of G r r}), which allow us to express $\bar{F}^r_r- \bar{F}^A_A$ as $-2 \bar{\lambda}$ and $\bar{G}^r_r$ as $\frac{\beta}{4\sqrt{\bar{\gamma}}}\;\bar{p}$, respectively. Note that, in the case of considering Lorentzian signature ($s=-1$), the charge presented in equation (\ref{ST Ash = ST ADM}) is equivalent to that of equation (\ref{STcharge}), since $\tilde{f}$ and $T$ are both arbitrary even functions.

\section{\textsf{Conclusion and outlook}}
In this paper, we have put forward novel boundary conditions for the Ashtekar-Barbero variables at spatial infinity within the framework of asymptotically flat spacetimes. These new boundary conditions are described by equation (\ref{New Ash BC}), with the parity conditions stated in the theorem of section (\ref{Section Explicit form}). We have examined these boundary conditions without resorting to the ADM expressions.

These boundary conditions satisfy the following consistency requirements: the symplectic structure is well-defined and the boundary conditions are preserved under hypersurface deformations. It turns out that by using the new parity conditions, the generators of the asymptotic symmetries are finite only for spacetime translations, but not for boosts and rotations. This issue also arises when working with ADM variables, but in \cite{Henneaux}, the authors provide a strategy to resolve it, which involves imposing faster fall-off conditions for the constraints, namely the additional conditions (\ref{LOT of constraints ADM}).
In this paper, we have imposed conditions (\ref{LOT radial Gauss}), \ref{LOT angular Gauss}, (\ref{LOT radial Diff}), (\ref{LOT angular Diff}), and (\ref{LOT Ham.}), with the hope that a similar strategy will eliminate the divergence in boundary terms obtained in terms of Ashtekar-Barbero variables. Contrary to expectations, this strategy did not render the surface terms finite, and in fact, the charges corresponding to boosts and rotations remain divergent. In section \ref{Section comparison with H-T}, we have analyzed the reason for this discrepancy and conclude that with the new boundary conditions, ADM variables and Ashtekar-Barbero variables cannot be considered canonically equivalent, and this distinction manifests itself in the boundary terms. 

Nonetheless, the significant and noteworthy accomplishment of the present work is that if we disregard boosts and rotations, the charge corresponding to translations is not only finite and integrable, but also incorporates supertranslations. In contrast to \cite{Thiemann, Campiglia}, where the charge corresponding to supertranslations vanishes and thus they are pure gauge, in this work, the supertranslation generators are not identically zero and therefore act non-trivially in the physical phase space. Thus, we have successfully achieved the objective outlined in the introduction, which is to associate standard canonical generators at spatial infinity with supertranslations in terms of Ashtekar-Barbero variables initially observed at null infinity.
The parity conditions in the theorem play a crucial role in the new boundary conditions, distinguishing themselves from the previously proposed conditions in \cite{Thiemann, Campiglia}. It has been duly noted that the odd parity of $W$ and the even parity of $\tilde{f}$ define the characteristics of the supertranslations. These parities, which are incompatible with the parity conditions laid out in \cite{Thiemann, Campiglia}, do not exist in that approach except when it comes to Poincar\'e translations.
\\
\\
This work has potential for extension and further exploration in multiple directions:
\\
\\
\begin{enumerate}
\item[1.] Our initial motivation for investigating new boundary conditions in the Ashtekar-Barbero framework is to incorporate techniques from LQG in order to establish a quantum theory. In future research, our objective is to construct quantum operators corresponding to the charges of supertranslations using holonomy and flux operators, and then investigate their quantum behaviors. 

\item[2.] As the charges associated with boosts and rotations are not finite under the proposed boundary conditions, our future work will focus on identifying alternative boundary conditions that overcome this limitation. We have observed that finiteness of the symplectic structure can be achieved through parity conditions, and therefore we must constrain ourselves to those parity conditions that yield a finite symplectic structure. One possible approach to relax these parity conditions is to adopt the framework of holographic renormalization to remove the divergences appearing in the symplectic structure, as proposed in \cite{Compere:2011ve}.

\item[3.] Once we have successfully accomplished the previous goal, our next aim is to determine boundary conditions that not only yield supertranslations at spatial infinity, but also incorporate superrotations \cite{Barnich:2009se}.
\end{enumerate}


\subsection*{\textsf{Acknowledgement}} 
The author expresses gratitude to John Joseph Marchetta for his encouragement and valuable discussions. The present work is funded by the National Natural Science Foundation of China (NSFC) under Grants no. 12275022 and no. 11875006.

\appendix

\section{\textsf{Variations of the leading order terms of the canonical variables}}\label{App. Preservation}
In this appendix, we proceed with the calculation of the variation of $\bar{F}^a_b$ and $\bar{G}^a_b$ under hypersurface deformations, as used in section \ref{Section Preservation}. We begin by examining the variations of the variables associated with the densitized triad $E^a_i$. It is important to note that the variation of $E^a_i$ under a hypersurface deformations can be obtained through a combination of equations (\ref{Var. of E under Guass Const.}), (\ref{Var. of E under diff Const.}), and (\ref{Var. of E under Ham Const.}). By utilizing the asymptotic expansion of $\delta E^a_i$, we can readily determine the variation of $\bar{f}^a_i$. Specifically, we have that $\delta E^r_i = r \sqrt{\bar{\gamma}} \delta \bar{f}^r_i + O(1)$ and $\delta E^A_i = \sqrt{\bar{\gamma}} \delta \bar{f}^A_i + O(r^{-1})$. It is worth mentioning that we assume that $\bar{\gamma}^a_i$ is not subject to variations under hypersurface deformations, i.e. $\delta \bar{\gamma}^a_i =0$. Once we have obtained the expression for $\delta \bar{f}^a_i$, we can determine the variation of $\delta \bar{F}^a_b$ as $\delta \bar{F}^a_b = \delta (\bar{f}^a_i \bar{\gamma}^i_b) = \bar{\gamma}^i_b (\delta \bar{f}^a_i)$. This result follows again from the fact that $\delta \bar{\gamma}_a^i =0$.

Using the above strategy and after a rather lengthy calculation, we have obtained all the radial and angular components of $\delta \bar{F}^a_b$ as follows.
\begin{align}\label{Var. F r r}
\delta \bar{F}^r_r &= \beta \left(\frac{\epsilon^{AB}}{\sqrt{\bar{\gamma}}} \bar{D}_A (b \bar{F}^r_B) - \frac{\epsilon^{AB}}{\sqrt{\bar{\gamma}}}\bar{\gamma}_{AC} b \bar{F}^C_B 
- b \bar{G}^A_A + \frac{(\beta^2 - s)}{\beta} b  \bar{k}^A_A - \frac{\epsilon^{AB}}{\sqrt{\bar{\gamma}}}(\partial_A b)\bar{F}^r_B \right)
+
2W + \bar{D}_A I^A + \mathcal{L}_Y \bar{F}^r_r \nonumber\\
&=
\beta \left(\frac{\epsilon^{AB}}{\sqrt{\bar{\gamma}}} b \bar{D}_A \bar{F}^r_B - \frac{\epsilon^{AB}}{\sqrt{\bar{\gamma}}}\bar{\gamma}_{AC} b \bar{F}^C_B 
- b \bar{G}^A_A + \frac{(\beta^2 - s)}{\beta} b  \bar{k}^A_A \right)
+
2W + \bar{D}_A I^A + \mathcal{L}_Y \bar{F}^r_r
\end{align}

\begin{align}\label{Var. F r D}
\delta \bar{F}^r_D =& \beta \left(\bar{\gamma}_{BD} \frac{\epsilon^{BA}}{\sqrt{\bar{\gamma}}} \left[
\bar{D}_A (b \bar{F}^r_r) 
-b \bar{\gamma}_{AC} \bar{F}^C_r
+ \bar{F}^C_A (\bar{D}_C b) 
+ (\bar{D}_A \tilde{f})\right]
+ b \; \bar{G}^A_r \bar{\gamma}_{AD} - \frac{(\beta^2 - s)}{\beta} b \; \bar{k}^r_D\right) \nonumber\\
& - \beta \frac{\epsilon^{AB}}{\sqrt{\bar{\gamma}}}\bar{\gamma}_{AD}\left(-b \bar{F}^r_B + (\partial_B b)\bar{F}^r_r\right) 
- \partial_D W + I_D + \mathcal{L}_Y \bar{F}^r_D + \frac{\epsilon^{AB}}{\sqrt{\bar{\gamma}}}\bar{\gamma}_{AD} \bar{\Lambda}^C \bar{\gamma}_{BC}\nonumber\\
=& \beta \left(\bar{\gamma}_{AD} \frac{\epsilon^{AB}}{\sqrt{\bar{\gamma}}} \left[
b \bar{D}_B \bar{F}^r_r 
-b (\bar{\gamma}_{BC} \bar{F}^C_r - \bar{F}^r_B)
+ \bar{F}^C_B (\bar{D}_C b) 
+ (\bar{D}_B \tilde{f})\right]
+ b \; \bar{G}^A_r \bar{\gamma}_{AD} - \frac{(\beta^2 - s)}{\beta} b \; \bar{k}^r_D\right) \nonumber\\
& - \partial_D W + I_D + \mathcal{L}_Y \bar{F}^r_D + \frac{\epsilon^{AB}}{\sqrt{\bar{\gamma}}}\bar{\gamma}_{AD} \bar{\Lambda}_B 
\end{align}

\begin{align}\label{Var. F B r}
\delta \bar{F}^B_r =& \beta \left[\frac{\epsilon^{AC} }{\sqrt{\bar{\gamma}}} \bar{D}_A (b\bar{F}^B_C) + \frac{\epsilon^{CB} }{\sqrt{\bar{\gamma}}}\bar{F}^A_C (\bar{D}_A b )  + \frac{\epsilon^{AB}}{\sqrt{\bar{\gamma}}} (\bar{D}_A \tilde{f})+ b \left(\bar{G}^r_A \bar{\gamma}^{AB} - \frac{(\beta^2 - s)}{\beta} \bar{k}^B_r \right)- \frac{\epsilon^{AC}}{\sqrt{\bar{\gamma}}}(\partial_A b) \bar{F}^B_C\right]\nonumber \\
&+ \mathcal{L}_Y \bar{F}^B_r - \frac{\epsilon^{BA}}{\sqrt{\bar{\gamma}}}\bar{\gamma}_{AC} \bar{\Lambda}^C \nonumber \\
=& 
\beta \left[\frac{\epsilon^{AC} }{\sqrt{\bar{\gamma}}}b \; \bar{D}_A \bar{F}^B_C + \frac{\epsilon^{CB} }{\sqrt{\bar{\gamma}}}\bar{F}^A_C (\bar{D}_A b )  + \frac{\epsilon^{AB}}{\sqrt{\bar{\gamma}}} (\bar{D}_A \tilde{f})+ b \left(\bar{G}^r_A \bar{\gamma}^{AB} - \frac{(\beta^2 - s)}{\beta} \bar{k}^B_r \right)\right]+ \mathcal{L}_Y \bar{F}^B_r - \frac{\epsilon^{BC}}{\sqrt{\bar{\gamma}}}\bar{\Lambda}_C
\end{align}

\begin{align} \label{Var. F B D}
\delta \bar{F}^B_D =& \beta\left[\frac{\bar{\gamma}_{CD}}{\sqrt{\bar{\gamma}}} \left(\epsilon^{CA} \bar{D}_A (b \bar{F}^B_r) - \epsilon^{CB} \bar{F}^A_r (\bar{D}_A b) - \epsilon^{CA} b \bar{F}^B_A \right)\right. \nonumber\\
&\hspace*{1cm} \left. + \sqrt{\bar{\gamma}} b\;(\bar{G}^C_A \bar{\gamma}^{AB} \bar{\gamma}_{CD} - \delta_D^B \bar{G}^b_b   )
+ \frac{(\beta^2 - s)}{\beta} \sqrt{\bar{\gamma}} b\;  (\bar{k}^b_b \delta_D^B - \bar{k}^B_D )\right]\nonumber\\
&- \beta \frac{\epsilon^{CA}}{\sqrt{\bar{\gamma}}}\bar{\gamma}_{CD} \left(-b \bar{F}^B_A + (\partial_A b)\bar{F}^B_r \right) + W \delta^B_D - \bar{D}_D I^B + \delta^B_D (\bar{D}_A I^A) + \mathcal{L}_Y \bar{F}^B_D + \frac{\epsilon^{AB}}{\sqrt{\bar{\gamma}}}\bar{\Lambda}^r \bar{\gamma}_{BD}\nonumber \\
=&
\beta\left[\frac{\bar{\gamma}_{CD}}{\sqrt{\bar{\gamma}}} \left(\epsilon^{CA} b \bar{D}_A \bar{F}^B_r - \epsilon^{CB} \bar{F}^A_r (\bar{D}_A b)\right) + b\;(\bar{G}^C_A \bar{\gamma}^{AB} \bar{\gamma}_{CD} - \delta_D^B \bar{G}^b_b)
+ \frac{(\beta^2 - s)}{\beta}  b\;  (\bar{k}^b_b \delta_D^B - \bar{k}^B_D )\right]\nonumber\\
&+ W \delta^B_D - \bar{D}_D I^B + \delta^B_D (\bar{D}_A I^A) + \mathcal{L}_Y \bar{F}^B_D + \frac{\epsilon^{BC}}{\sqrt{\bar{\gamma}}}\bar{\Lambda}^r \bar{\gamma}_{CD}
\end{align}

A similar approach should be employed for $A_a^i$. Note that the variation of $A_a^i$ under hypersurface deformations can be determined by combining (\ref{Var. of A under Guass Const.}), (\ref{Var. of A under diff Const.}), and (\ref{Var. of A under Ham Const.}). By using the asymptotic expansion of $\delta A_a^i$, the variation of $\bar{g}^a_i$ can be easily deduced as follows: $\delta A_r^i = \frac{1}{r^2} \delta \bar{g}_r^i + O(r^{-3})$ and $\delta A_A^i = \frac{1}{r} \delta \bar{g}_A^i + O(r^{-2})$. Once $\delta \bar{g}_a^i$ is obtained, we can find $\delta \bar{G}^a_b$ using the equation $\delta \bar{G}^a_b = \delta (\bar{g}_b^i \bar{\gamma}_i^a) = \bar{\gamma}_i^a (\delta \bar{g}_b^i)$, since it is known that $\delta \bar{\gamma}^a_i = 0$.


Using the above strategy and after a tedious calculation, we have obtained all the radial and angular components of $\delta \bar{G}^a_b$ as follows.

\begin{align}\label{Var. G r r}
\delta \bar{G}^r_r =& -\beta \frac{b}{\sqrt{\bar{\gamma}}}\epsilon^{AC} \bar{\gamma}_{BC} \left( \bar{D}_A \bar{G}^B_r +  \bar{G}^B_A  \right) + \frac{\beta^2 -s}{2 \sqrt{\bar{\gamma}}} \bar{D}_A (b\; \epsilon^{AB} [\bar{\gamma}_{CB} \bar{k}^C_r + \bar{k}^r_B]) - \beta \frac{\epsilon^{AC}}{\sqrt{\bar{\gamma}}}(\partial_A b)\bar{\gamma}_{BC} \bar{G}^B_r \nonumber\\
&+ \mathcal{L}_Y \bar{G}^r_r  + \bar{\Lambda}^r
\end{align}

\begin{align}
\delta \bar{G}^B_r =& -\beta \frac{b}{\sqrt{\bar{\gamma}}}\epsilon^{AB} \left(-\bar{D}_A\bar{G}^r_r + \bar{\gamma}_{AC} \bar{G}^C_r - \bar{G}^r_A\right)\nonumber \\
& + \frac{\beta^2 -s}{2 \sqrt{\bar{\gamma}}} \left( -2 \bar{D}_A (b \;\epsilon^{AB} \bar{k}^r_r)- \bar{D}_A (b \;\epsilon^{AB}\bar{k}^C_C) -2 b\; \epsilon^{BC} \bar{\gamma}_{AC}\bar{k}^A_r + \bar{D}_A (b \;\epsilon^{AC} \bar{k}^B_C - b \;\epsilon^{BC} \bar{k}^A_C)\right) \nonumber\\
& -\beta  \frac{\epsilon^{BA}}{\sqrt{\bar{\gamma}}} \left((\partial_A b) \bar{G}^r_r -b \; \bar{\gamma}_{AC} \bar{G}^C_r\right) + \mathcal{L}_Y \bar{G}^B_r + \bar{\Lambda}^B
\end{align}

\begin{align}
\delta \bar{G}^r_B =& -\beta \frac{b}{\sqrt{\bar{\gamma}}} \epsilon^{DA} \left(\bar{\gamma}_{BD} \bar{G}^r_A + \bar{\gamma}_{DC} \bar{D}_B \bar{G}^C_A - \bar{\gamma}_{CD} \bar{D}_A \bar{G}^C_B \right)\nonumber\\
& + \frac{\beta^2 -s}{2 \sqrt{\bar{\gamma}}} \left(\bar{\gamma}_{BC} \bar{D}_A(b\;\epsilon^{AC} \bar{k}^E_E) - 2b \;\epsilon^{CA} \bar{\gamma}_{AB} \bar{k}^r_C - \bar{\gamma}_{BD}\bar{D}_A (b \;\epsilon^{CD}\bar{k}^A_C + b\;\epsilon^{CA}\bar{k}^D_C) \right)\nonumber\\
& -\beta  \frac{\epsilon^{AD}}{\sqrt{\bar{\gamma}}} (\partial_A b) \bar{\gamma}_{CD} \bar{G}^B_C  + \mathcal{L}_Y \bar{G}^r_B - \bar{D}_B \bar{\Lambda}^r + \bar{\gamma}_{AB}\bar{\Lambda}^A
\end{align}

\begin{align}
\delta \bar{G}^D_B =& \beta \frac{b}{\sqrt{\bar{\gamma}}}\left(\epsilon^{CD}\bar{\gamma}_{BC}\bar{G}^r_r + \epsilon^{CD}\bar{\gamma}_{AC} \bar{D}_B \bar{G}^A_r + \epsilon^{DA} \bar{D}_B\bar{G}^r_A -\epsilon^{DA}\bar{\gamma}_{BC}\bar{G}^C_A - \epsilon^{DA}\bar{D}_A \bar{G}^r_B \right)\nonumber\\
& + \frac{\beta^2 -s}{2 \sqrt{\bar{\gamma}}} \left(-2 \epsilon^{AD} b \bar{\gamma}_{AB} \bar{k}^r_r - \epsilon^{AD} b \bar{\gamma}_{AB} \bar{k}^C_C + \epsilon^{AD} b \bar{\gamma}_{BC} \bar{k}^C_A + \bar{\gamma}_{BC} \bar{D}_A (\epsilon^{CA} b \bar{k}^D_r) - \epsilon^{CA} b \bar{\gamma}_{AB} \bar{k}^D_C  \right. \nonumber\\
& \hspace{1.8cm} \left.- \bar{\gamma}_{BC} \bar{D}_A (\epsilon^{CD} b \bar{k}^A_r) + \epsilon^{AE} b \bar{\gamma}_{EC} \bar{k}^C_A \delta^D_B - \bar{\gamma}_{BC} \bar{D}_A (\epsilon^{AD} b \bar{k}^C_r) + \delta^D_B \bar{D}_A (\epsilon^{EA}b [\bar{k}^r_E - \bar{\gamma}_{EC}\bar{k}^C_r]) \right) \nonumber\\
& - \beta \frac{\epsilon^{DA}}{\sqrt{\bar{\gamma}}} \left((\partial_A b)\bar{G}^r_B - b \bar{G}^C_B \bar{\gamma}_{AC} \right)
+ \mathcal{L}_Y \bar{G}^D_B - \delta^D_B \bar{\Lambda}^r - \bar{D}_B \bar{\Lambda}^D 
\end{align}

}
\end{document}